\documentclass[a4paper, preprint, authoryear]{elsarticle}

\usepackage{amssymb}
\usepackage[nodots]{numcompress}
\usepackage{url}
\usepackage{amsmath}
\usepackage{tabularx}
\usepackage{multirow}
\usepackage{rotate}
\usepackage{paralist}
\usepackage{color, colortbl}
\usepackage{flushend}

\definecolor{Gray}{gray}{0.9}

\newcommand{\specialcell}[2][c]{%
  \begin{tabular}[#1]{@{}l@{}}#2\end{tabular}}

\journal{Information Processing and Management}

\begin{document}

\begin{frontmatter}
  \title{Internet Advertising: An Interplay among Advertisers, Online
    Publishers, Ad Exchanges and Web Users}
\author{Shuai Yuan, Ahmad Zainal Abidin, Marc Sloan, Jun Wang}
\address{Department of Computer Science, University College London}

\begin{abstract}
  Internet advertising, aka Web advertising or online advertising, is
  a fast growing business. It has already proved to be significantly
  important in digital economics. For example, it is vitally important
  for both web search engines and online content providers and
  publishers because web advertising provides them with major sources
  of revenue. Its presence is increasingly important for the whole
  media industry due to the influence of the Web. For advertisers, it
  is a smarter alternative to traditional marketing media such as TVs
  and newspapers. As the web evolves and data collection continues,
  the design of methods for more targeted, interactive, and friendly advertising may
  have a major impact on the way our digital economy evolves, and to
  aid societal development.

  Towards this goal mathematically well-grounded Computational
  Advertising methods are becoming necessary and will continue to
  develop as a fundamental tool towards the Web.  As a vibrant new
  discipline, Internet advertising requires effort from different
  research domains including Information Retrieval, Machine Learning,
  Data Mining and Analytic, Statistics, Economics, and even
  Psychology to predict and understand user behaviours.  In this
  paper, we provide a comprehensive survey on Internet advertising,
  discussing and classifying the research issues, identifying the
  recent technologies, and suggesting its future directions.  To have
  a comprehensive picture, we first start with a brief history,
  introduction, and classification of the industry and present a
  schematic view of the new advertising ecosystem. We then introduce
  four major participants, namely advertisers, online publishers, ad
  exchanges and web users; and through analysing and discussing the
  major research problems and existing solutions from their
  perspectives respectively, we discover and aggregate the fundamental
  problems that characterise the newly-formed research field and capture its
  potential future prospects.

\end{abstract}

\begin{keyword}
Internet advertising \sep online advertising \sep computational advertising
\sep web advertising  \sep search engine marketing \sep display
advertising \sep contextual advertising \sep sponsored search \sep
user behaviour targeting \sep demand side platform \sep supply side
platform \sep ad exchange \sep ad network
\end{keyword}

\end{frontmatter}


\section{Introduction}
\label{sec-intro} 

Advertising is a marketing message that attracts potential customers to purchase a product or to subscribe to a service. In addition, it is a way to establish a brand image through repeated presence of an advertisement (ad) associated with the brand in the media. Traditionally, television, radio, newspaper, magazines, and billboards are among the major channels that place ads. The advancement of the Internet and the World Wide Web (WWW) enables users to seek information online. Using the Internet and the WWW, users are able to express their information requests, navigate specific websites and perform e-commerce transactions. Major search engines have been continuing improving their retrieval services and users' browsing experience by providing relevant results. The Internet and the WWW are therefore a natural choice for advertisers to widen their strategy in reaching potential customers among Web users. 

This phenomenon provides an opportunity for the search engine to be a strategic platform for advertisers to place their ads on the Web, with the view that a proportion of those who are online and seeking specific products or services may click the ads. Currently, Web advertising is seen as complementing and we believe soon it will possibly dominate existing media as the preferred medium for placing ads, because one of the major advantages that it has over traditional advertising media is that the former is more targeted. A user expressing his or her information need in the form of a query, e.g. \textit{car rental}, is likely to respond to ads relevant to that query listed along with the organic search results. In comparison, ads in the newspaper have been pre-selected for display even before readers pick up their copies, and are less targeted and uniform for every reader. In addition, it is also not easy to measure the success of the advertising due to the lack of an effective feedback mechanism in the conventional media.

The revenue from Internet advertising shows a positive trend. It was reported that search engines' revenues from search advertising (sponsored search) in 2006 reached \$9.4 billion, a remarkable increase of approximately \$3.65 billion from the revenues earned in the previous year \citep{newcomb:marketing_07}. A survey by the Interactive Advertising Bureau shows that the revenue of Internet advertising in the US for the first half of 2009 reached over \$10.9 billion \citep{IAB2009}. In the first quarter of 2011, Internet advertising revenues reach \$7.3 billion \citep{IAB2011}. With this upward trend in revenue, the future of Internet advertising looks promising.  Please note that the term ``Internet advertising'' used throughout this paper also refers to Web advertising, online advertising or computational advertising.  In subsequent sections, these words are interchangeably used.

\subsection{A Brief History of Internet Advertising} 

Internet advertising has been around for over a decade. The \textit{sponsored search} paradigm was created in 1998 by Bill Gross of Idealab with the founding of Goto.com, which became Overture in October 2001, then acquired by Yahoo! in 2003 and is now Yahoo! Search Marketing \citep{Jansen2007Sponsored}. Meanwhile, Google started its own service \textit{AdWords} using \textit{Generalized Second Price Auction} (GSP) in February 2002 and added quality-based bidding in May 2002 \citep{Karp2008Google}. In 2007, Yahoo! Search Marketing added quality-based bidding as well \citep{Dreller2010Brief}. It is worth mentioning that Google paid 2.7 million shares to Yahoo! to solve the patent dispute in 2004 \citep{Google2004Dispute}, for the technology that matches ads with search results in sponsored search.
Web search has now become a necessary part of daily life, vastly reducing the difficulty and time that was once associated with satisfying an information need. Sponsored search allows advertisers to buy certain keywords to promote their business when users use such a search engine, and contributes greatly to its free service. 

In 1998 the history of \textit{contextual advertising} began. Oingo, started by Gilad Elbaz and Adam Weissman, developed a proprietary search algorithm based on word meanings and built upon an underlying lexicon called WordNet. Google acquired Oingo in April 2003 and renamed the system \textit{AdSense} \citep{Karp2008Google}. Later, Yahoo! Publish Network, Microsoft adCenter and Advertising.com Sponsored Listings amongst others were created to offer similar services \citep{Kenny2011Contextual}. Nowadays the contextual advertising platforms evolved to adapt to a richer media environment, such as video, audio and mobile networks with geographical information. These platforms allowed publishers to sell blocks of space on their web pages, video clips and applications to make money. Usually such services are called an \textit{advertising network} or a \textit{display network}, that are not necessarily run by search engines and can consist of a huge number of individual publishers and advertisers. 

One can also consider sponsored search ads as a form of contextual ad that matches with very simple context -- queries, which has been emphasized due to its early development, large market volume and warm research attentions. In this paper we will continue to use this categorization and take sponsored search as an example to illustrate common challenges for Internet advertising. 

Around 2005, new platforms focusing on real-time buying and selling impressions were created, like ADSDAQ, AdECN, DoubleClick Advertising Exchange, adBrite, and Right Media Exchange, that are now known as \textit{ad exchanges}. Unlike traditional ad networks, these ad exchanges aggregate multiple ad networks together to balance the demand and supply in marketplaces. Individual publishers and advertising networks can both benefit from participating in such businesses: publishers sell impressions to advertisers who are interested in associated user profiles and context; advertisers, on the other hand, could also get in touch with more publishers for better matching. At the same time, other similar platforms emerged \citep{Graham2010Brief} like \textit{demand side platform} (DSP) and \textit{supply side platform} (SSP), which are discussed further in Section-\ref{sec-classification}. However real-time bidding and multiple ad networks aggregation do not change the nature of such marketplaces (where buying and selling impressions happen). For simplicity we always use the term ``ad exchange'' in this paper to better represent the characteristics of platforms where trading happens.

\subsection{Characteristics of Internet Advertising}

In this paper, the Internet advertising types we focus on are sponsored search, contextual ads and branding ads. Since the types of advertising that we are going to discuss require computation and a principled way of finding the \textit{best match} between a given user in a given context and available ads, they are also referred to as computational advertising. In recent years, Internet advertising has been seen as a fast growing, scientific research sub-discipline involving established research areas such as microeconomics, information retrieval, statistical modelling, machine learning and recommender systems. 

The \textit{best match}, however, is not limited to the `relevance' from the traditional informational retrieval research sense, but also includes the \textit{best revenue} from the economic perspective. For instance, in the context of sponsored search, the challenge for search engines is to find and display the best ads from advertisers which suit user\rq{}s interest (relevance) as well as generating as much revenue as possible. The dilemma of balancing relevance and revenue is discussed in detail in Section-\ref{sec-exchange}. The \textit{best match} challenge naturally leads to the heavy dependence on computing power and algorithm design, especially given that billions of queries are received and handled everyday \citep{comscore2010search}, and many more webpages are visited and rendered (with associated contextual and branding ads).

Another significant difference between Internet advertising and traditional advertising business is its effective advertising cost. From the advertisers' point of view, the cost to advertise online is variable by choosing different pricing models, among which the most popular are cost-per-click (CPC), cost-per-mille (CPM) and cost-per-acquisition (CPA). For example, the CPC pricing model only charges the advertiser whenever an ad is clicked, which reflects the interest of the user. This is based on the effective targeting ability, which in turn leads back to the \textit{best match} challenge.

Even if the pricing model is chosen, the final cost is variable due to the competition in auctions. The auctions are carried out every time the ads need to be displayed and also takes into account the quality score of historical performance and landing pages for the ads. This encourages advertisers to improve their campaigns in all aspects rather than increase the bids solely. The auctions used in Internet advertising are discussed in Section-\ref{sec-exchange}.

By contrast, the cost of using traditional advertising media is usually fixed and determined or negotiated before the ads are deployed. In addition, traditional advertising media do not support real-time bidding in which an advertiser is able to specify a bid for a given user and context in real-time.

\begin{figure}[t!]	
	\centering
	\includegraphics[width=.7\textwidth]{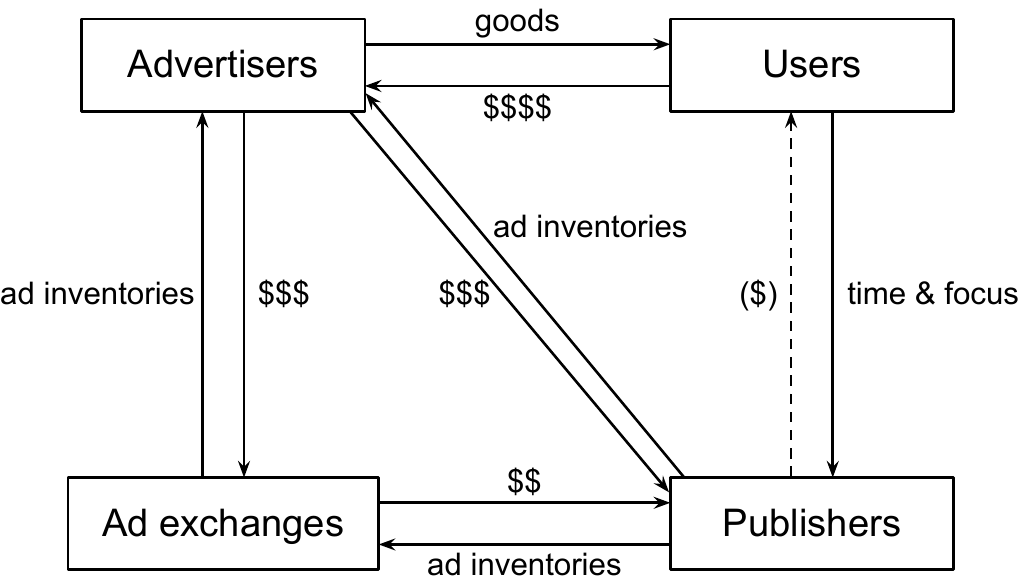}
	\caption{The simplified ecosystem of Internet advertising. Advertisers spend budget to buy ad inventories from ad exchanges and publishers; ad exchanges serve as matchers for ads and inventories; publishers provide valuable information to satisfy and keep visitors; users read ads and purchase goods from the advertisers. Note that normally users would not receive cash from publishers.}
	\label{fig-simplified-ecosystem}
\end{figure}

\subsection{A Schematic View of the Internet Advertising Ecosystem} 

First we analyse the sustainability of the Internet advertising business by presenting the general view of the ecosystem in Figure-\ref{fig-simplified-ecosystem}.  There are four main participants in Internet advertising: \textit{ad exchange}, \textit{advertiser}, \textit{publisher} and \textit{user}. 

If we take an analogy from the economic perspective, ad inventories are traded based on the force of demand and supply. An advertiser demands his ads to be displayed, whereas a publisher sells his ad inventories to gain revenue. In the case of sponsored search, a search engine acts as a publisher who has reserved space for ads on the search result page, whereas in contextual advertising, a content publisher reserves some space for ads. Referring to Figure-\ref{fig-simplified-ecosystem}, the descriptions of the participants are as follows: 

An \textit{ad exchange} is normally an advertising service providing the mechanism that enables advertisers to promote their products to targeted groups of users. The ad network/exchange acts as auctioneer, selling keywords to advertisers. Examples of advertising services include Google AdWords\footnote{\url{http://www.google.com/adwords} (last visited 02/06/2011)} and Yahoo! Sponsored Search\footnote{\url{http://searchmarketing.yahoo.com} (last visited 02/06/2011)}. The match between (i) keywords (ads) and query terms; (ii) keywords and webpage contents and (iii) keywords and user historical data are processed in ad exchanges. An ad exchange also manages contract negotiation between advertisers and content publishers that wish to sell ad spaces. The advertising marketplaces are becoming more complicated with the emergence of DSP, SSP and data exchange (DX) as well as the expanding of traditional business giants like Visa and MasterCard, who are trying to employ credit card data to target Internet ads \citep{Steel2011Using}.

An \textit{advertiser} requires spaces to place its marketing messages (i.e. ads) on search result pages in sponsored search and on webpage reserved spaces in the context of contextual advertising. \citet{joachims:ctr_ir05} argue that for each ad, its position and the total number of ads on the page have a significant influence on its click-through rate (CTR). In the sponsored search framework, bid prices and the relevancies between bid phrases and user queries influence the awarded slot position. However, the bids placed by other advertisers on similar keywords are unknown, whether each bid will end up winning a slot is uncertain. Ads displayed at higher positions are more likely to be clicked on, therefore, advertisers typically compete to bid for keywords that they believe to be relevant to user queries to increase the chances that their ads will be placed at top positions. 

A content \textit{publisher} hosts websites that may reserve spaces for display advertisements. The publisher usually employs the brokering services of an advertising platform such as Google's DoubleClick, AOL's Advertising.com and Microsoft Media Network. For large publishers, normal practice is to sell only remnant inventory through an ad exchange, with the other inventory being negotiated directly with advertisers. Smaller publishers normally sell all of their advertising spaces through ad exchanges. We note that a search engine acts as a publisher in the sponsored search case where conceptually its role is not much different from that of a content publisher. 

A \textit{user} issues ad-hoc topics to express his or her information needs. In organic search, the relevance between a search topic (query) and documents on the Web is used to retrieve relevant documents. However, in search-based advertising, ads are not retrieved purely based on relevance. The match between the search topics and the advertisers' keywords, the bid prices and CTRs for the keywords are among factors of deciding which ads are eventually given ad slots, although the exact method is unique from one search engine to another. The selected ads will be displayed alongside organic search results, for a pay-per-click model, the advertisers will be charged only if there are clicks on their displayed ads. 

In the figure, we also present the conceptual view of cash flow in the advertising ecosystem, where the cycle and volume of cash flow between participants can be seen. Note that the amount of cash flow from advertisers to ad exchanges, ad exchanges to publishers, publishers to users, and users to advertisers are not necessarily equivalent nor in any proportion. Besides, the value passed from publishers to users is normally not in money, but rather information or service. In traditional advertising media such as newspapers and magazines, users need to purchase the newspaper or magazines in order to view the contents and ads. In Internet advertising, publishers (except those that require a subscription) usually allow their web content to be visited for free, provided a transfer of the benefit it receives from advertising revenue.  

In ad exchanges, the interactions between advertisers and the ad exchange are to bid keywords for ads (e.g. using Google AdWords) to be listed in a search result page or publisher's website at the auctioned price; or to negotiate contracts for branding ads at a fixed price (e.g. using Yahoo! My Display Ads\footnote{\url{http://advertisingcentral.yahoo.com/smallbusiness/mydisplayads} (last visited 13/12/2011)}). The interactions between an ad exchange and publishers are the placements of relevant ads for the spaces that publishers offer at auctioned or fixed prices. For spaces sold at auctioned prices, publishers use solutions (e.g. Google AdSense\footnote{\url{http://www.google.com/adsense} (last visited 13/12/2011)}) to bolster the revenue for displaying relevant ads. Through a search result page or publisher's website, a user's search query or browsing context are used by the ad exchange to return the best relevant ads. The process of bidding and targeting is discussed in detail in Section-\ref{sec-exchange}.

The main objectives of this paper are to provide a comprehensive survey on Internet advertising, discuss research issues, identify the state-of-the-art technologies and suggest its future directions. The organization of the paper is as follows: First we give classifications of elements in advertising business in Section-\ref{sec-classification}; Research issues and methodologies from the view of ad exchanges are discussed in Section-\ref{sec-exchange}, advertisers in Section-\ref{sec-advertiser}, publishers in Section-\ref{sec-publisher}, and users in Section-\ref{sec-user}.  More specifically, the state-of-the-art research work on auction design, relevance calculation and revenue management for ad exchanges are firstly presented and compared; they are followed by recent research for advertisers, particularly keyword discovery and selection, bid optimisation and experiment design; which leads onto a discussion on revenue management and maximisation for publishers; and finally the research regarding users (or targets) of online advertising, more specifically click-through models, behavioural targeting and user privacy concerns, will be given.  Concluding remarks and future directions of computational advertising are given in Section-\ref{sec-future}.  The popular terminology is enumerated in \ref{sec-terminology}.
 
\section{Classification}
\label{sec-classification}
Before proceeding to research issues and methodologies, it is necessary to clarify some concepts and classify the elements that comprise the Internet advertising business.

\subsection{Players}
\begin{figure}[t!]
	\centering
	\includegraphics[width=\textwidth]{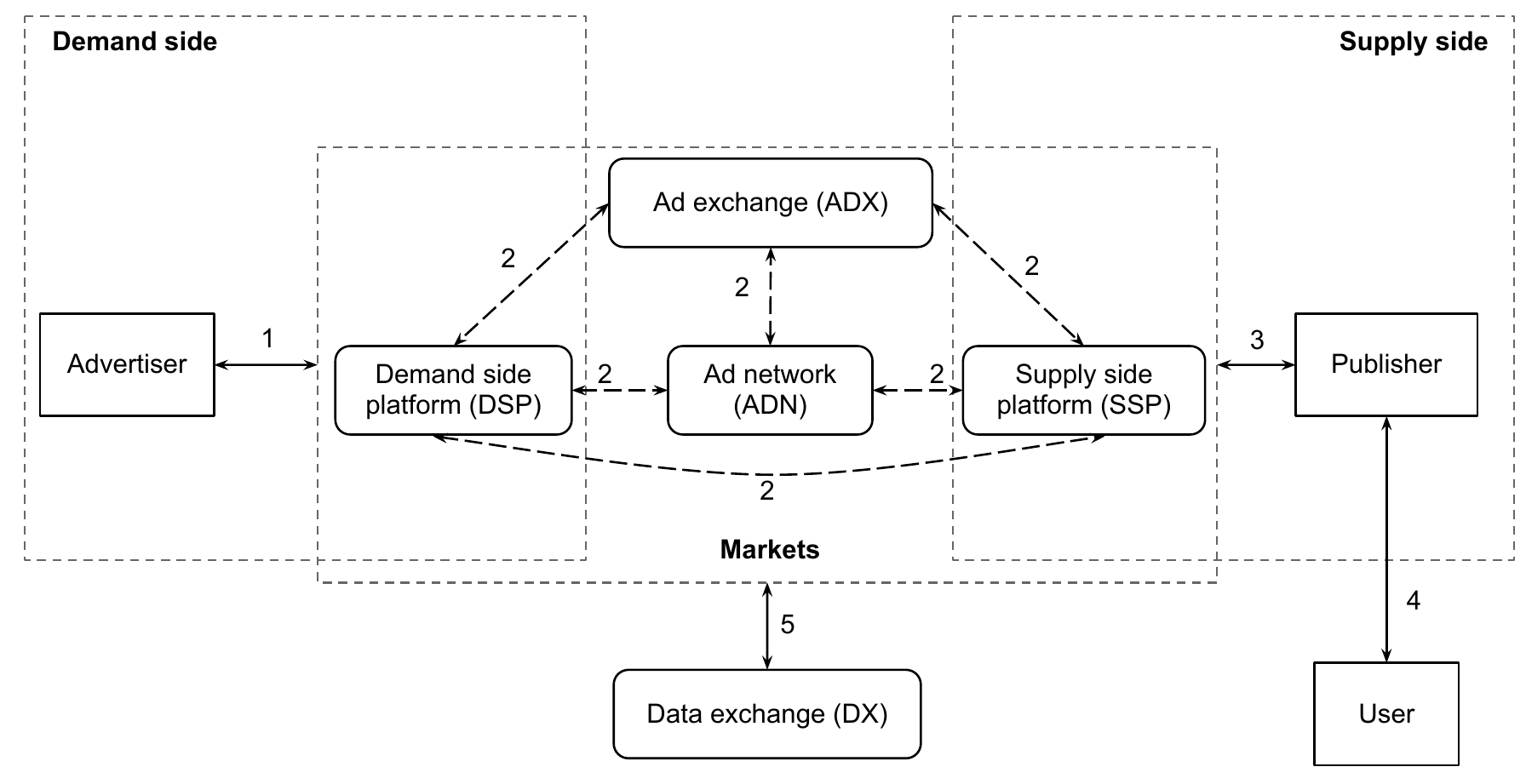}
	\caption{The various players of Internet advertising and the trading process: 1) The advertiser creates campaigns in markets 2) The markets can trade campaigns and impressions to balance the demand and supply for better efficiency 3) The publisher registers impressions with the markets 4) The user issues queries or visits webpages 5) The markets can query data exchanges for user profiles in real-time bidding}
	\label{fig-players}
\end{figure}

Traditionally there are four major types of player in the Internet advertising business as introduced before: \textit{advertisers}, \textit{publishers}, \textit{ad networks} and \textit{users}. However in recent years, with the rapid development of the industry and growing revenue, increasing numbers of companies are engaging with the business by providing new tools and platforms which makes them unique and valuable to the traditional players, as shown in Figure-\ref{fig-players}:
\begin{compactitem}
	\item \textbf{Demand side platforms} (DSP) serve advertisers or ad agencies by bidding for their campaigns in multiple ad networks automatically;

	\item \textbf{Supply side platforms} (SSP) serve publishers by registering their inventories (impressions) in multiple ad networks and accepting the most beneficial ads automatically;

	\item \textbf{Ad exchanges} (ADX) combine multiple ad networks together \citep{muthukrishnan2009ad}. When publishers request ads with a given context to serve users, the ADX contacts candidate Ad Networks (ADN) in real-time for a wider selection of relevant ads;

	\item \textbf{Data exchanges} (DX), sometimes called Data Management Platforms (DMP), serves DSP, SSP and ADX by providing user historical data (usually in real-time) for better matching.
\end{compactitem}

The emergence of DSP, SSP, ADX and DX is a result of the fact that there are thousands of ad networks available on the Internet, which can act as a barrier for advertisers as well as publishers when getting into the online advertising business. Advertisers have to create and maintain campaigns frequently for better coverage, and analyse data across many platforms for a better impact. Publishers have to register with and compare several ad networks carefully to achieve optimal revenue. The ADX came as an aggregated marketplace of multiple ad networks to help alleviate such problems. Advertisers can create their campaigns and set desired targeting only once and analyse the performance data stream in a single place, and publishers can register with ADX and collect the optimal profit without any manual interference.

The ADX could be split into two categories, namely DSP and SSP, for their different emphasis on customers. The DSP works as the agency of advertisers by bidding and tracking in selected ad networks, the SSP works as the agency of publishers by selling impressions and selecting optimal bids. However the key idea behind these platforms is the same: they are trying to create a uniform marketplace for customers; on the one hand to reduce human labour and on the other hand to balance demand and supply in various small markets for better economic efficiency.

Due to the opportunities and profit in such business, the borderline between these platforms is becoming less tangible. In our work we choose to use the term ``ad exchange'' to describe the marketplace where impression trading happens.

The DX collects user data and sells it anonymously to DSP, SSP, ADX and sometimes advertisers directly in real-time bidding (RTB) for better matching between ads and users. This technology is usually referred to as \textit{behaviour targeting}. Intuitively, if a user's past data shows interest in advertisers' products or services, then advertisers have a higher chance of securing a transaction by displaying their ads, which results in higher bidding for the impression. Initially the DX was a component of other platforms, but now more individual DXs are operating alongside analysing and tracking services.
\begin{figure}[t!]	
	\centering
	\includegraphics[width=\textwidth]{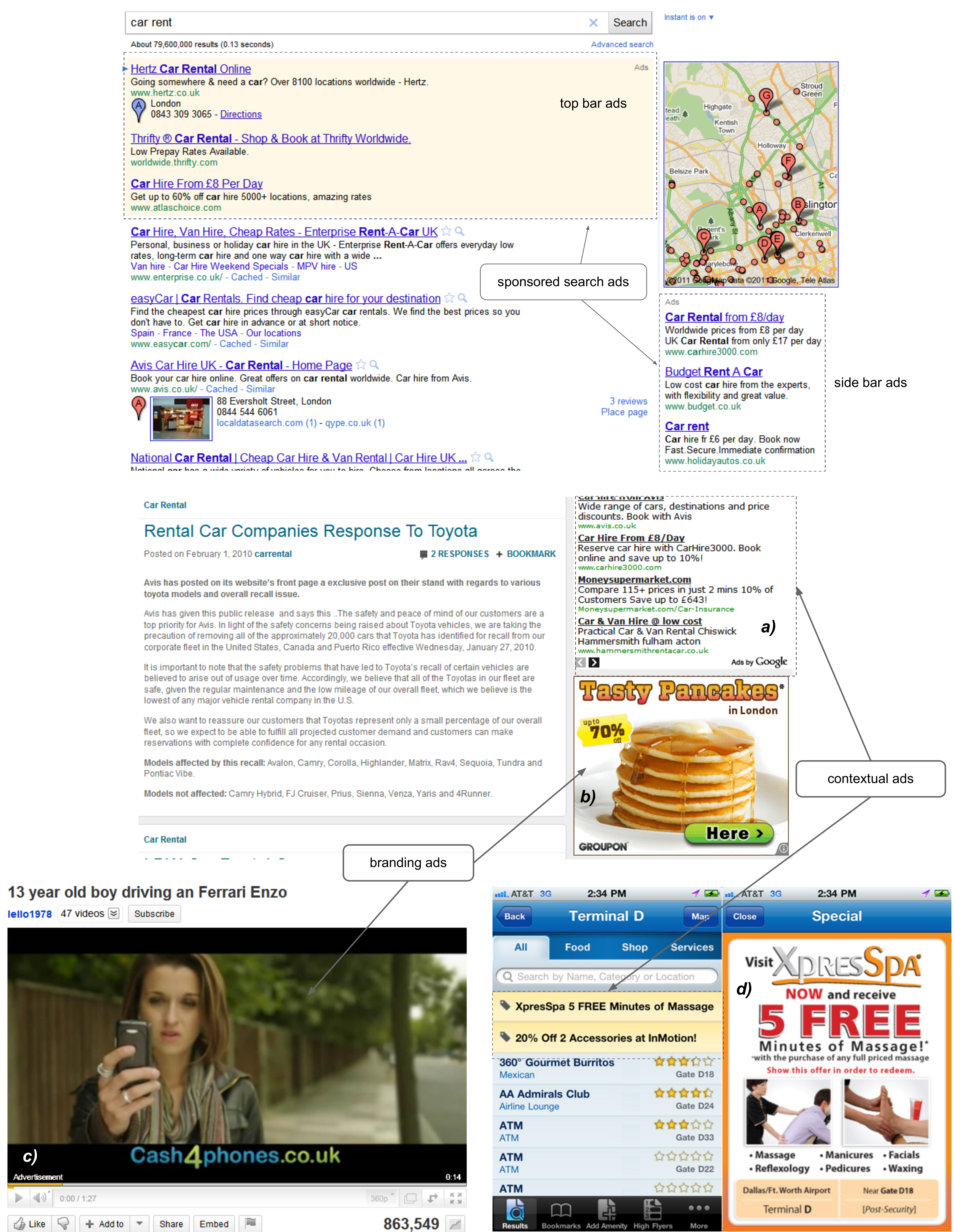}
	\caption{Common types of ads on the Internet. At the top are sponsored search ads relevant to the query issued by the user. Section \textbf{a)} and \textbf{d)} are contextual ads while \textbf{b)} and \textbf{c)} are branding ads (not likely behaviour targeting ads since cookies were off), which are not clearly relevant with the context.}
	\label{fig-ads}
\end{figure} 

\subsection{Ads}

Internet ads can be categorised into the following types and their variants:
\begin{compactitem}
	\item \textbf{Sponsored search ad}. Search engines act as publishers by displaying ads alongside search results. Advertisers can buy keywords that exactly or broadly match with queries submitted by visitors. \textit{Paid listing}, \textit{Paid search} or \textit{Paid inclusion} are known variants.

	\item \textbf{Branding ad}. Advertisers buy impressions of webpages from publishers and have their ads displayed to all visitors. This is commonly seen in over-the-counter (OTC) contracts and branding campaigns, where targeting is so broad that context can be ignored. For example, in Figure-\ref{fig-ads} the ads of `pancake' and `cash for phone' (marked as \textbf{b)} and \textbf{c)} respectively) are clearly not relevant to the context \textit{car rental} or \textit{13 year old boy driving a Ferrari Enzo}.

	\item \textbf{Contextual ad}. Instead of displaying the same ads to everyone, different ads are shown with regard to the geography, language, device and other characteristics of visitors, to maximize the utilization of advertising opportunities. From Figure-\ref{fig-ads} we can see that section \textbf{a)} is related to the browsing context, whilst section \textbf{d)} is relevant based on the user's geographical location. Note that although behaviour targeting ads are not represented in Figure-\ref{fig-ads} we consider them a variant of contextual ads, where the weight of content relevance is much less than that of user relevance.
\end{compactitem}

Also note that in some literature \textit{display ad} is used to refer to both branding ads and contextual ads. A detailed comparison is presented in Table-\ref{tbl-ads}.

\newcolumntype{g}{>{\columncolor{Gray}}l}
\begin{table}
\scriptsize{	
	\centering
	\caption{Comparison of different types of ads.} 
	\begin{tabularx}{\textwidth}{|g|X|X|X|} 
	\hline \rowcolor{Gray}
	& Branding ads & Sponsored search ads & Contextual ads \\ \hline

	Seller & Large content providers & Search engines & All websites \\ \hline
	Buyer & Large companies & \multicolumn{2}{c|}{ All companies and individuals } \\ \hline
	Contract & Long-term, large scale & \multicolumn{2}{c|}{ Flexible } \\ \hline
	Targeting & None & Queries and Personally Identifiable Information (PII) & Context and PII \\ \hline
	Ad assets & Fixed slots on webpages & Top and side banners in search result page & Flexible \\ \hline
	\end{tabularx}
	\label{tbl-ads}
}
\end{table}

\subsection{Delivery}

The delivery of impressions can be divided into:
\begin{compactitem}
	\item \textbf{On the spot}, mainly used for transparent markets (TM) where auctions are held when visitors show up. The ads from winning advertisers are displayed immediately. 

	\item \textbf{Forward contracts}, mainly used for OTC contracts and are also observed in transparent markets. The advertiser purchases impressions in advance for some future time period instead of the current time period, e.g. Q1 next year, to make sure that the advertising opportunity is guaranteed for their business plans. 
\end{compactitem}

\subsection{Trading Places}

Generally, advertising opportunities are traded either \textbf{over-the-counter} (OTC), i.e. by private contract, or in \textbf{transparent markets} (TM) by some auction or reservation mechanism. The contracts agreed by advertisers and publishers, or between ad networks, are long term and of large volume; on the contrary, if advertising opportunities are traded in transparent markets, the trading units are usually small (e.g. 1k impressions), although the total number of impressions for the advertising campaign could be huge.

\subsection{Competition Methods} 

\begin{figure}[t!]	
	\centering
	\includegraphics[width=\textwidth]{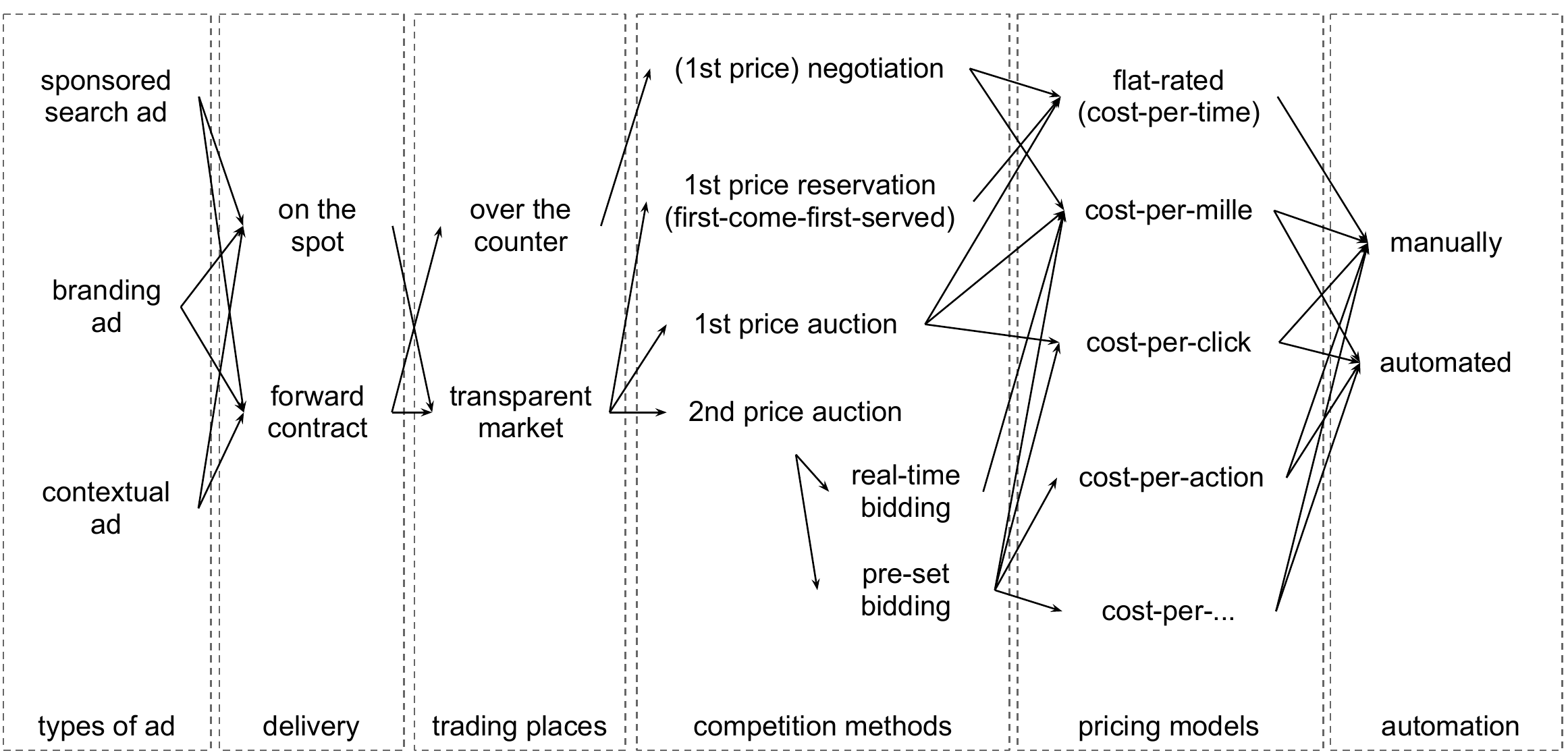}
	\caption{The trading route of Internet advertising business from an advertiser's perspective. The flexibility of choice between each section makes the associated companies highly specialized, thus there is no provider for all functionalities in one place. However, this could change in the future when the market becomes more consistent and standardized.}
	\label{fig-trading}
\end{figure}

There will always be competition in trading and there are several ways to facilitate competition and generate winners as shown in Figure-\ref{fig-trading}: 

\begin{compactitem}
	\item \textbf{1st price negotiation}, normally operated by a human. For example, a publisher reviews contracts submitted by advertisers and contacts the preferred one.

	\item \textbf{1st price auction}, created in 1996 by Open Text and then Goto.com in 1998 for their cost-per-click programs, gradually abandoned after the invention of Google AdWords in 2000. Some ad network startups now propose the idea of ad futures (delivering impressions in the future but not marketable), with which advertisers place 1st price bids if they do not want to offer an outright buy order \citep{constantin2009online}.

	\item Some ad networks choose \textbf{1st price reservation} to deal with advertising opportunities in future on a first-come-first-serve basis, instead of auctions. This case is similar with \textit{1st price negotiation} except that no negotiation is required and the trading could happen in \textit{transparent markets}.

	\item \textbf{2nd price auction}, or generalized second price auction (GSP) \citep{Edelman2007Internet} is the most popular technique used in today\rq{}s ad networks. Instead of paying for the bid offered, the advertiser pays the price calculated from the next highest bid and quality scores. For example, if advertiser A places a bid of \$10 with a quality score of 5, advertiser B places a bid of \$8 with a quality score of 8, and they both target exactly the same audience, then advertiser B wins because $\$8\times8 > \$10\times5$, and the actual price paid will be $\$10\times5/8=\$6.25$.

	Additionally, advertisers and publishers can choose between \textit{pre-set bidding} (PSB) or \textit{real-time bidding} (RTB) if they use GSP. With PSB the advertisers set the campaign, desired target, and bids before auctions; however, with RTB advertisers can bid each time before an impression is delivered, adjusting their bids according to user profiles and other factors.
\end{compactitem}

\subsection{Pricing Models}

The two sides of supply and demand must agree on a pricing model as well, i.e. how much to pay for a unit good. Some popular models used are:

\begin{compactitem}
	\item \textbf{Flat-rated} (or cost-per-time, CPT) and \textbf{cost-per-mille} (CPM) are commonly used when delivery of impressions is in the future. The former indicates that the cost (of some ad slot) will be constant (for some time) regardless of actual traffic (number of impressions) delivered by publishers. The latter takes actual traffic into account, however the CPM price is fixed due to the fact that no auction is held against other competitors.

	\item In spot markets the preferred pricing models are \textbf{cost-per-mille}, \textbf{cost-per-click}, \textbf{cost-per-action}, \textbf{cost-per-lead}, \textbf{cost-per-view}, and \textbf{cost-per-complete-view}.
\end{compactitem}

Additionally, if the advertiser chooses to use RTB, to our best extent of knowledge the only possible pricing model at present is CPM (as in 12/2011). The total cost receives greater variance due to the fact that the advertiser has to place the bid every time an impression is delivered. As for ad exchanges and publishers, they tend to receive higher revenue since advertisers are willing to bid more, given that every impression is better matched with visitors.

Regardless of the choice of pricing models, in Internet advertising,  the goods traded are always \textit{impressions}, even if the advertisers choose to use CPC, CPA, or others. By providing various pricing models in ad networks, more advertisers are attracted as they are alleviated from the burden of calculating the number of impressions needed to get a click or a conversion and thus how much they should bid. The ad networks handle the calculation and relevance matching, they have enough data to solve the problem as well as allowing them more control in their auctions, e.g. the quality score.

\subsection{Automation}

The whole process of creating and competing in Internet advertising trading can be either \textbf{manual} or \textbf{automated}. The manual operation with OTC trading is a natural part of the initial negotiation. Small advertisers and publishers using PSB tend to do campaign management or ad slots management manually to reduce the cost. With the various tools provided by ad networks, they can set up their campaign, specify targets and bids, analyse return-on-investment (ROI) or track revenue without many challenges. 

However, for RTB advertisers, the job is almost impossible to complete manually due to the high volume and speed of incoming requests for placing bids for every impression, which requires analysing the context, user profile and other data. Therefore, automated systems are employed in RTB platforms, enabling advertisers to give precise bids very quickly with the help of machine learning algorithms.

\begin{figure}[t!]	
	\centering
	\includegraphics[width=\textwidth]{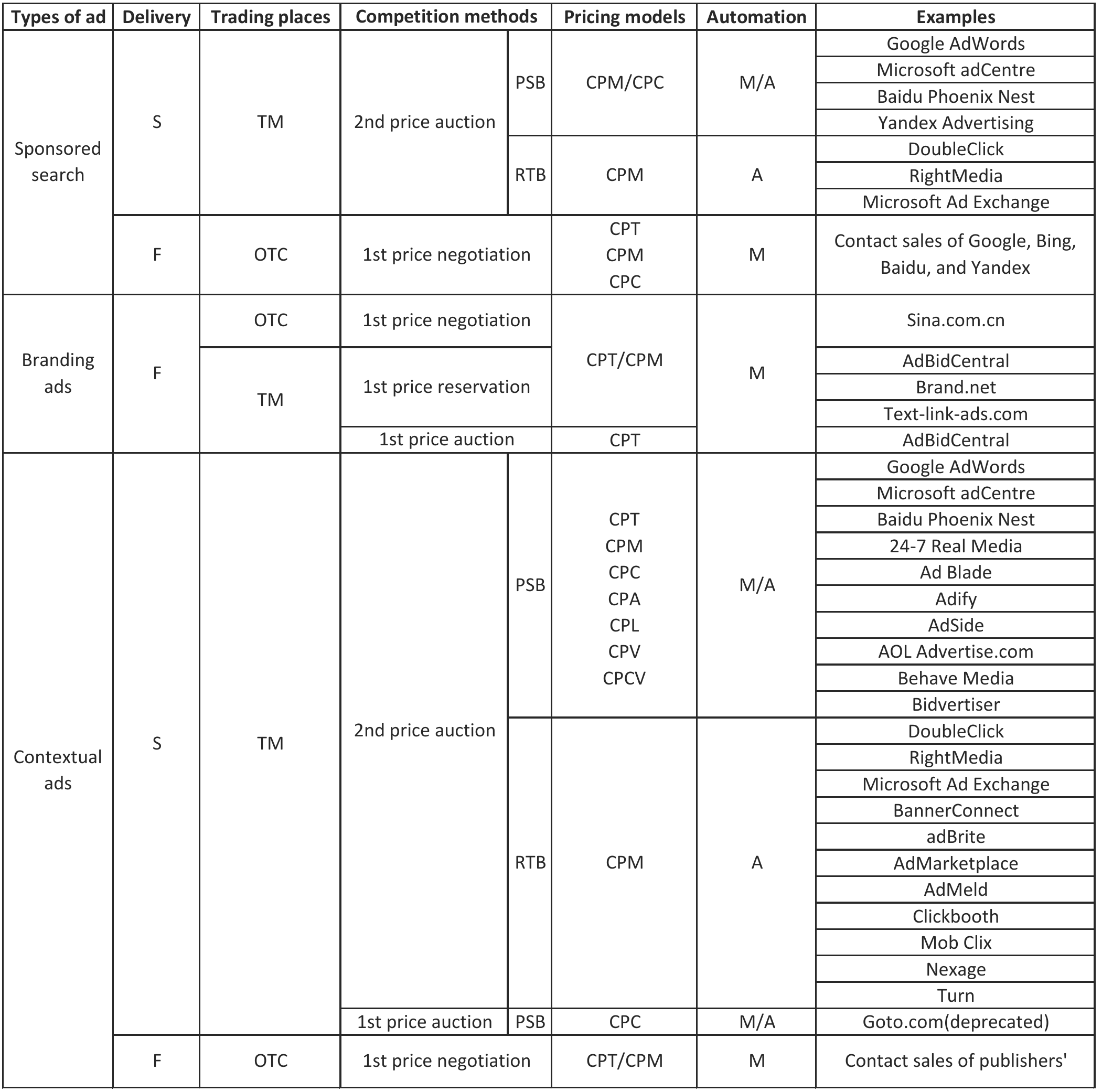}
	\caption{Examples of different trading methods. \textbf{S} stands for \textit{On the spot} and \textbf{F} for \textit{Forward contracts}. \textbf{M} stands for \textit{Manual} and \textbf{A} for \textit{Automated}.}
	\label{fig-examples}
\end{figure}

\subsection{Examples}

Here we list some typical examples, as shown in Figure-\ref{fig-examples}, in the Internet advertising business and analyse them regarding the four properties listed above:
\begin{compactitem}
	\item Sina.com.cn\footnote{\url{http://emarketing.sina.com.cn/} (last visited 13/12/2011)}. Mainly uses OTC contracts agreed by negotiation, using flat-rated or CPM as the pricing model. The trading is mostly completed manually.

	\item Break.com\footnote{\url{http://www.breakmedia.com/for-advertisers} (last visited 13/12/2011)}. Instead of solely relying on OTC contracts, they also sell their (remnant) impressions through ad networks \citep{roels2009dynamic}. Therefore, the problem of trade-off between fulfilling contracts and selling through ad networks is raised.

	\item Text-link-ads.com, AdBidCentral.com, and Brand.net are ad networks offering flat-rated pricing models to advertisers, with 1st price reservation. Additionally AdBidCentral.com also provides 1st price auction if the advertiser does not want to place an outright buy order.

	\item Google AdWords is the biggest ad network offering CPM, CPC and CPA with GSP.

	\item adBrite.com is an ad exchange offering \textit{cost-per-view}(CPV), which deploys a fullscreen ad before showing the hosting page.

	\item Doubleclick.com\footnote{\url{http://www.google.com/doubleclick/} (last visited 13/12/2011)}, Rightmedia.com, and Microsoft Advertisng Exchange\footnote{\url{http://advertising.microsoft.com/exchange} (last visited 13/12/2011)} are probably the most famous ad exchanges (operated by Google, Yahoo!, and Microsoft respectively) offering RTB at present.
\end{compactitem}

Having understood the Internet advertising ecosystem as a while, we are now ready to discuss the major research problems and their existing solutions for each of the key players, namely advertisers, online publishers, ad exchanges and web users in the following sections.

\section{The Perspective of an Ad Exchange}
\label{sec-exchange}

As discussed in Section-\ref{sec-classification}, in this paper the \textit{ad exchange} (ADX) is considered a uniform marketplace for publishers to sell ad inventories, and for advertisers to buy impressions and clicks. Such concepts have existed since the 1990s, big content providers such as media broadcasters created ad slots on their web pages in the form of banners, side bars, foot bars, blocks embedded in text etc, and sold these slots to advertisers with offline, long-term contracts. 

The name \textit{ad exchange} well defines the common characteristics of all advertising platforms including individual publishers' contracts, sponsored search and ad networks, whilst summarising the major activities involved: selling and buying ad slots. We believe that this technology will be mature enough to take over all selling and buying from other platforms and have all ad slots and their prices listed and changing in real-time similarly to stock markets. The ADX has a strong chance of becoming the only interface between publishers and advertisers. 

The basic ad elements used in online advertising are common, such as a headline, creative (could be text, pictures, or even video clips), a URL and a landing page. It's possible that the landing page can use a different URL, however there is often the restriction of limiting it to the same domain. For example, an advertiser can deploy an ad with a URL {\scriptsize \textit{www.WEBSITE.com/promote} } (which is shown to users) while the actual landing page is {\scriptsize \textit{www.WEBSITE.com/COMPLICATED-PATH} } (which is hidden for ease of use).

In addition, advertisers need to provide the bidding phrases, matching criteria (exact or broad match) and the budget (max CPC and daily budget). Due to the similarity of these features, advertisers can easily publish ads to a search network(\textit{sponsored search}) and display network at the same time.

Advertisers are also able to publish rich media ads on web pages like pictures or animations as long as they are supported by users' browsers. Moreover, with the rapid growth of Internet streaming services like Youtube and the maturity of hand-held smart devices like smart phones, pads and even smart cameras, advertisers are now able to publish ads whenever a user watches a video clip, searches for restaurants in a specific region with a mobile phone or even enters or leaves a building, as shown in Figure-\ref{fig-ads}. 

However, no matter which media is chosen, how the ads look and on what screen they're displayed, the essential challenges for ad exchanges remain the same: constructing a strong and simple auction model, balancing relevance and revenue optimization and dealing effectively with textual information. In this section we will discuss these challenges separately. Note these challenges are different from those in real-time bidding focused scenarios \citep{muthukrishnan2009ad}.

\subsection{Auction Models Construction}

The first challenge for ADX is creating a solid auction model to support the flow of the advertising eco-system. There is a lot of research on different auction models, including \cite{Muthukrishnan2008Internet, aggarwal2008theory, Varian2007Position, Edelman2007Internet}, and many related aspects have been evaluated and compared in \citet{Ghose2007Empirical}. Nowadays the \textit{generalized second price auction} (GSP) is the most adopted model in ADX. The GSP auction is a non-truthful auction model for multiple items, similar to a \textit{Vickrey-Clarke-Groves auction} (VCG) but without truthfulness. The process is defined in \citet{Edelman2007Internet}: for a given keyword, there are $N$ slots and $K$ bidders. Each slot has a probability of being clicked of $a_{i}$. We can assume that ads in top slots have a larger probability of being clicked (due to rank bias, see Section-\ref{sec-user}), so: 
\[
a_{0}\geq a_{1}...\geq a_{N}. 
\]

The value per click to the bidder $k$ is $s_{k}$. Bidders are risk-neutral, and bidder $k$'s payoff from being in position $i$ is equal to $a_{i} s_{k}- p_{k}$, where $p_{k}$ is the payment to the search engine. 

Suppose at some time $t$ a search engine user enters a given keyword, and for bidder $k$ the submitted bid for this keyword was $b_{k}$; if bidder $k$ did not submit a bid then $b_{k}=0$. Let $b(j)$ and $g(j)$ denote the bid and identity of the $j$-th highest bidder respectively. Note that if several bidders submit the same bid, they are ordered randomly. The model then allocates the top position to the bidder with the highest bid $g(1)$, the second position to $g(2)$ and so on, down to position $\min(N, K)$. Note that each bidder gets at most one slot.

If the search engine user clicks on a bidder's link, the bidder will be charged the \textit{next} highest bid. Therefore the bidder $g(i)$'s total payment is \[ p(i)=a_{i} b(i+1), i\in \{1,\ldots,\min(N,K)\}, \] where their payoff is \[ a_{i} (s_{i}-b(i+1)).  \]

If there are at least as many slots as bidders $(N\geq K)$ then the last bidder's payment $p(K)$ is zero. However, in realistic search engines using GSP, the charge is often slightly higher, e.g. Google charges an extra \$0.01 for each click. Therefore, the last bidder pays \$0.01 every time the user clicks the link.

\begin{figure}
	\centering
	\includegraphics[width=.4 \textwidth]{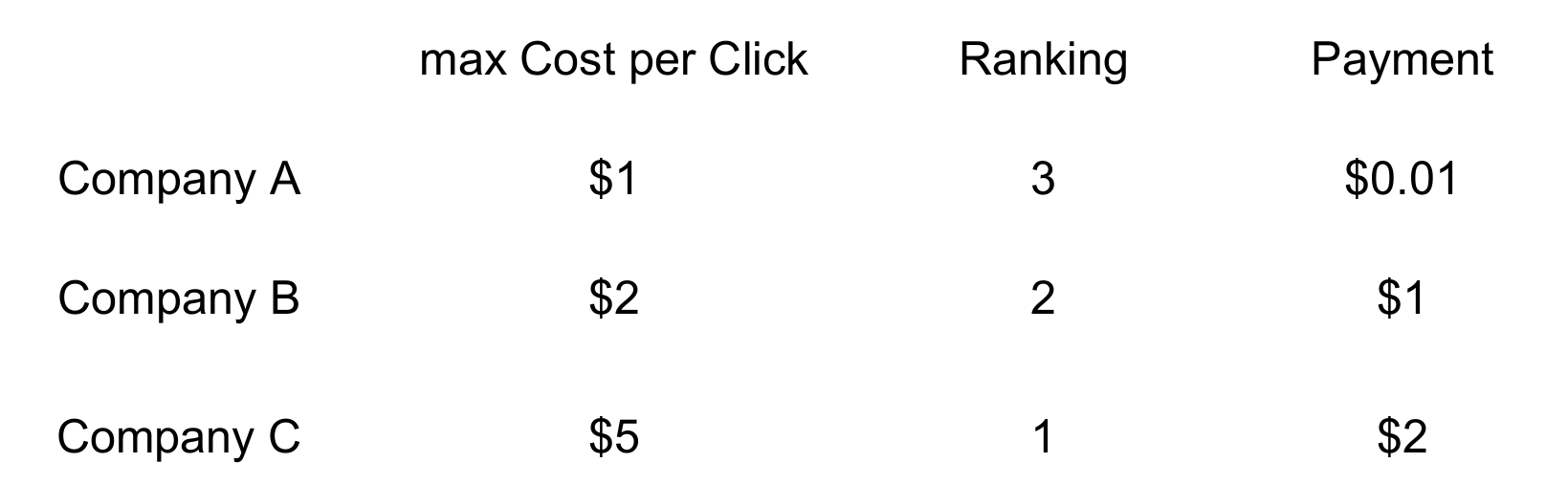}
	\includegraphics[width=.4 \textwidth]{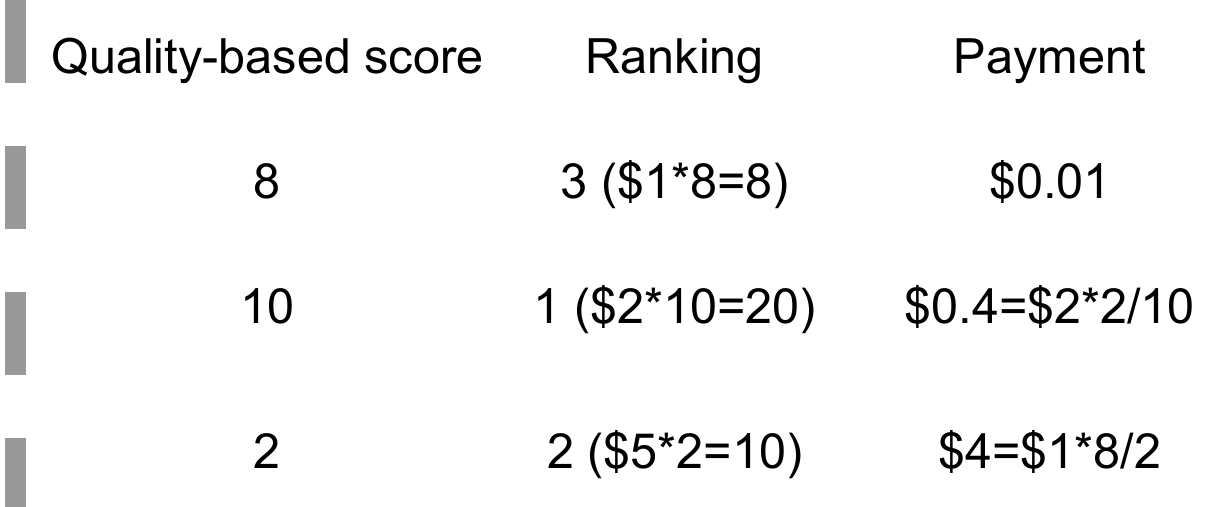}
	\caption{The bidding budget and the resulting ranking for companies A, B and C (left) and the change of rank and payment combined with the quality-based score (right).}
	\label{fig-bidding}
\end{figure}

For example, company A, B and C are in the car rental business and they are trying to attract users to their websites who search for ``car rent'' in the Google search engine. Thus, they create their advertisement campaigns using Google AdWords, choosing ``car rent'' as the keywords, and submit their own ads. Depending on the profit they could make from a user renting a car, company A, B and C choose the bidding budget (maximum Cost per Click) separately as \$1, \$2, and \$5. Suppose that the three companies are the only ones competing for the keywords ``car rent'' in Google AdWords and that there are exactly three ad slots for sponsored search. Then the results will be similar to Figure-\ref{fig-bidding}.

As mentioned before, a VCG auction is similar to the GSP auction but at the same time having the characteristic of truthfulness. The bidding and ranking part remains the same, however the VCG auction model charges bidders the negative externalities they impose on the other bidders, rather than simply the next highest price. Therefore the payment for bidder $g(i)$ is 
\[
p(i)=(a_{i}-a_{i+1})b(i+1)+p(i+1). 
\]
and the last bidder winning a slot pays zero if $N\geq K$, otherwise $a_{N}b(i+1)$.

We can speculate about the reason why popular search engines choose GSP rather than VCG. Firstly, VCG has the drawbacks of low seller revenue, vulnerability to collusion and other fraudulent auction behaviour as described in \citet{ausubel2006lovely}; secondly, \textit{negative externality} is not easy to explain and may increase the complexity of the marketing platform, which is supposed to be user friendly; thirdly, it would be expensive to implement a new model, test and then switch the whole system, especially for established companies like Google. We consider the GSP auction model to be a good balance between constructing a concrete platform and winning a big crowd of customers, and with years worth of improvement, it's no longer challenging for a new advertiser to understand all of the associated concepts and operate on the platforms.

At present, sponsored search advertising service providers like Google and Yahoo! use quality-based scores to encourage advertisers to create better content. Given a bidding phrase and a landing page, the quality-based score is computed according to the relevance between keywords and the web page, with supplements like the historical data of a user's account and the CTR of the page. Therefore, let $q_{0}$,\ldots,$q_{k}$ denote the quality score for $k$ bidders. The bid for bidder $i$ for a given keyword in the above scenario becomes $b'(i)=b(i)q(i)$. The payment to the search engine becomes 
\[
p'(i)=\frac{q(i+1)}{q(i)p(i+1)},
\]

Therefore, if advertisers are able to get a higher quality-based score, they can reduce their cost. However, one should note that ADX also use quality-based scores to set limitations, meaning that advertisers may have to bid more to get in the competition, even if there is no competitor. For example, if the minimum allowed quality score $\times$ bid is 10 in Google AdWords, company A now has to bid at least $\$1\times\frac{10}{8}=\$1.25$ for their ad to be displayed.

\begin{table}[!t]
\small{
	\centering
	\caption{Research on relevance optimization.}
	\label{tbl-relevance-methods}
	\begin{tabularx}{\textwidth}{|X|X|X|X|} \hline
\rowcolor{Gray}
Goal & Author(s) & Method & Data \\ \hline

\multirow{4}{*}{Predict CTR} & \citet{Richardson2007Predicting} & Logistic regression & 10k advertisers, 1m ads, 500k keywords \\ \cline{2-4}

 & \citet{ciaramita2008online} & Perceptron & 11k ad blocks \\ \hline

\multirow{5}{*}{\specialcell{Determine ads \\ delivery}} & \citet{Jin2007swc} & SVM, logistic regression & 100k pages, 200 judgments for each category \\ \cline{2-4}

 & \citet{broder2008swing} & SVM & 1k+ pages, 642 queries, 29k judgements \\ \hline

\multirow{5}{*}{Improve relevance} & \citet{Neto2005Impedance} & Vector space model, Bayesian networks & 1.7k advertisers, 94k ads, 68k keywords\\ \cline{2-4}

& \citet{hillard2010improving} & Max-entropy,  adaBoost decision tree stump, gradient boosting decision trees & 22k judgements \\ \hline

Improve relevance and revenue & \citet{Radlinski2008Optimizing} & Regression SVM & 10m queries, each has 3 ads \\ \hline

\multirow{5}{*}{\specialcell{Find ads for \\ long-tail queries}}& \citet{broder2008search} & Vector space model & 10m queries, 9k query-ad pair and judgements\\ \cline{2-4}

 & \citet{Broder2009oe} & Vector space model & 100m queries, 3 ads for each query, 3.5k judgements\\ \hline

Predict bounce rates & \citet{Sculley2009Predicting} & Regression SVM, parallel logistic regression & 10m data points\\ \hline

	\end{tabularx}
}
\end{table}

\subsection{Relevance Optimization}
\label{subsec-relevance}

When an ad slot needs to be filled on a webpage, the ad exchange decides:
\begin{compactitem}
	\item How many ads to show.
	\item Which ads to show.
	\item Where to show them.
\end{compactitem}

The maximum number of ads to be selected for a slot is usually fixed. For example, there are 3 slots at the top (north) and 8 slots on the right (east) on Google search result pages. Naturally, ad networks/exchanges try to maximize the revenue based on the probability that an ad will be clicked (CTR) as well as the payment of the ad (bid) \citep{Richardson2007Predicting}, however if relevance is valued less, then a consequence is that the user satisfaction can be reduced, resulting in less users coming back and eventual loss of revenue. On the other hand, adding highly relevant ads to a hosting page will improve the user experience by providing manually selected, directly related information, and particularly for the sponsored search scenario, sometimes the ads serve as straightforward solutions to the user's demands.

By matching ads to context, their relevance or similarity can be measured using the Vector Space Model, which is adopted by many researchers as shown in Table-\ref{tbl-relevance-methods}. In \citet{Neto2005Impedance} the problem is defined as: let $w_{iq}$ by the weight associated with term $t_i$ in the query $q$, and $w_{ij}$ the weight associated with term $t_i$ in the document $d_j$. Then $\vec{q} = (w_{1q}, w_{2q}, \ldots, w_{iq}, \ldots, w_{nq})$ and $\vec{d_j} = (w_{1j}, w_{2j}, \ldots, w_{ij}, \ldots, w{_nj})$ are the weighted vectors used to represent the query $q$ and the document $d_j$ . 

When using the vector space model these features are usually introduced:
\begin{compactitem}
	\item Word and character overlap (unigram, bigram etc.).
	\item Categorisation.
	\item Title and its length, creative and URL.
	\item Other minor factors, like dollar sign and commercial words.
\end{compactitem}

Some researchers \citep{Wang2002Understanding, Weideman2004Ethical} have reduced the relevance problem to the binary case, which determines if any ads should be shown at all when a query is submitted; \citet{broder2008swing} calls this a \textit{swing problem}. They proposed a machine learning approach formulating the decision into a binary classification problem. Instead of scoring every ad, the system predicts whether a set of ads is relevant enough to the given query. The authors used a Support Vector Machine (SVM) to learn the classification model with a wide range of features, including query/ad similarity and topical cohesiveness. The empirical experiments show that with a greedy feature selection strategy, the method is capable of achieving at least 69\% accuracy on both content match and sponsored search data sets. 

\citet{Jin2007swc} proposed the problem of whether ads should be displayed in sensitive content pages, such as news of a natural disaster, and built a sensitive content taxonomy with a hierarchical classifier using SVM and Logistic Regression. Their experiments show that the accuracy of the classification is around 80\%. Whilst their research is based around contextual advertising, the idea could also be interesting in the sponsored search scenario.

When queries and bidding phrases are relatively short, it's challenging for ADX to make good matches. \citet{Neto2005Impedance} proposed to generate an augmented representation of the target page by means of a Bayesian model built over several additional Web pages. \citet{Radlinski2008Optimizing} proposed an online query expansion algorithm of two stages: the offline processing module pre-computes query expansions for a large number of queries, and then builds an inverted index from the expanded query features. The inverted index maps features of expanded queries into the queries they characterize. In the online phase, when a new query arrives that has not been previously expanded in the offline phase, it is mapped onto related previously-seen queries using the inverted index, by matching the features of the incoming query against the features of previously expanded queries. \citet{Broder2009oe} extended the system for better rare queries matching.

Later, \citet{broder2008search} proposed a method for both augmenting queries and ads. Queries and ads are presented in three distinct spaces of different features: unigrams, classes and phrases extracted using a proprietary variant of Altavista's Prisma refinement tool. The query and ad vectors are evaluated by the vector space model, and the empirical experiments achieved 65\% precision at 100\% recall. The paper also reported an interesting observation that the quality of feature selection based on DF and  TF-IDF is essentially the same.

There are also researchers trying to use user click-through data to select relevant ads \citep{Richardson2007Predicting, ciaramita2008online, Sculley2009Predicting, hillard2010improving}. The user behaviour (clicking on the ad, moving to another task immediately after clicking and so on) is used as the judgement of relevance of the ad to the given query. These methods have the potential of reaching higher user satisfaction compared to traditional IR methods, however acquiring sufficient user behaviour data is very expensive and limits the research work in this direction (see Section-\ref{sec-user} for more information on this topic).

\begin{table}[!t]
\small{	
	\centering
	\caption{Research on semantic advertising.}
	\label{tbl-semantic-methods}
	\begin{tabularx}{\textwidth}{|X|X|X|X|} \hline
\rowcolor{Gray}
Goal & Author(s) & Method & Data \\ \hline

Find ads keywords on web pages & \citet{Yih2006Finding} & Logistic regression & 1109 pages\\ \hline

Match contextual ads & \citet{broder2007semantic} & Vector space model & 105 sampled from 20m webpages \\ \hline

\multirow{6}{*}{\specialcell{Insert ads to \\ video scenes}} & \citet{Li2005Real} & Consistent line detection, consistent less-informative region detection & 3 hours baseball documentary video\\ \cline{2-4}

 & \citet{wan2003real} & The Hough Transform-based line-mark detection & 7 full football games video\\ \hline

Match and insert ads to video scenes & \citet{Mei2007VideoSense} & Nonlinear 0-1 integer programming & 32 video clips selected from 14k candidates \\ \hline

\multirow{2}{*}{\specialcell{Match ads with \\ user information \\ on mobile devices}} & \citet{Mahmoud2006Provisioning} & Mobile agents provisioning & N/A \\ \cline{2-4}

 & \citet{Idwan2008Enhancing} & N/A & 200 customers and 30 shops\\ \hline

Provide comparison shopping using geographic information & \citet{Mahmoud2006Havana} & Mobile agents provisioning & N/A \\ \hline

	\end{tabularx}
}
\end{table}

\subsubsection{Semantic Advertising}\label{subsec-semantic}

The relevance problem can be easily extended when more elements are introduced, for instance in the display ads category, there are contextual ads (text ads), multimedia ads (deployed in videos) and mobile ads (embedded in apps or mobile-version hosting pages). The matching problem for contextual ads can be simplified if keywords can be extracted from the target web page. \citet{Yih2006Finding} pursued this line of thought and proposed a system extracting keywords from web pages using a number of features, like the term frequency (TF) of each potential keyword, inverse document frequency (IDF), presence in meta-data and how often the term occurs in MSN search query logs. The system showed around 25\% improvement compared to the KEA extractor\footnote{\url{http://www.nzdl.org/Kea/} (last visited 13/12/2011)} proposed by \citep{frank1999domain}.

However, a web page can provide more useful information than a query. \citet{Neto2005Impedance} proposed ten strategies and evaluated their effectiveness individually, with the restriction of only accessing the text of hosting pages as well as keywords and text for the ads. They reported that the \textit{matching strategies} achieved around 60\% improvement of precision compared to trivial vector-based strategies, and the \textit{impedance coupling strategies} managed an additional 50\% improvement. The matching strategies tried to match ads directly to given web pages, while impedance coupling strategies recognised that there is a vocabulary impedance problem among ads and Web pages and attempted to solve the problem by expanding the Web pages and the ads with new terms. However, the methodology is considered challenging for pricing as well as time consuming by \citep{Yih2006Finding}, for its cosine similarity computation between each hosting page and a bundle of words associated with each ad.

\citet{broder2007semantic} tried to classify web pages and ads into a 6000 nodes commercial advertising taxonomy and determined their distance by cosine similarity metrics. The results were complemented by syntactic matching between a bag of keywords from the web page and that from ads. They reported around 25\% improvement by combining the two parts mentioned above, then using syntactic matching alone.

Multimedia ads based on context have taken a step forward as hosting pages have become more video and audio oriented. Instead of calculating the relevance between text and ads, multimedia ads, particularly video ads on sharing sites like Youtube\footnote{\url{http://www.youtube.com}}, address the challenge of understanding multimedia contents. Revver (which is no longer in operation) selected one relevant ad (either a static picture or a video clip) and displayed it at the end of a video, sharing advertisement revenue with the video author. \citet{wan2003real} proposed a similar approach of detecting the goal-mouth appearance of soccer video in real-time. \citet{Li2005Real} proposed a system to detect baseball video scenes for commercial ads insertion. The proper timing was detected using a Hough Transform to calculate the consistent existence of simple background objects in the video for a period of time. The proper location was identified by finding flat regions in a uniform colour without any edge information in the video. Additionally, authors set algorithmic rules to make sure that the critical information of the video was not blocked by the ads, and that the ads were always stable, clear and viewable on the stationary background of the video. 

\citet{Mei2007VideoSense} proposed a system to understand video content and insert appropriate and relevant ads whenever the clip was being played. The ads would be inserted at the positions with highest discontinuity and lowest attractiveness, whilst the overall global and local relevance was maximized. First, candidate ads were selected according to text tags and descriptions of the video. Then a variant of Best-First Model Merging proposed by \citep{Zhao01Video} was adopted to detect video discontinuity, and user attention was estimated by a model from \citep{Ma2002A}; and the ad's insertion point was determined by these two factors. The author also discussed an optimization for online video ads insertion.

Ads on mobile devices has a long history \citep{mohamed2003wireless}, especially after Short Message Service (SMS) became popular. There is plenty of research about advertising using the traditional SMS channel \citep{Tsang2004Consumer, Barwise2002Permissionbased, haghirian2005increasing}. However, as the technology has advanced, researchers have become more concerned about increasing the effectiveness of employing modern technology like contextual analysis \citep{Mahmoud2006Provisioning, mohamed2003wireless} and using geographic location\citep{Mahmoud2006Havana, aalto2004bluetooth, Wang2005Locationbased, mohamed2003wireless, Idwan2008Enhancing}.

\begin{table}
\small{	
	\centering	
	\caption{Research for revenue optimization.}
	\label{tbl-revenue-methods}
	\begin{tabularx}{\textwidth}{|X|X|X|X|} \hline
\rowcolor{Gray}
Goal & Author(s) & Method & Data \\ \hline

Improve revenue & \citet{Mehta2005AdWords} & A trade-off-revealing family of LP & N/A \\ \hline

Improve revenue & \citet{bala:ad_ranking_webDB08} & A ranking function considering mutual influences & Simulation of 2m rounds\\ \hline

Balance between relevance and revenue & \citet{Zhu2009Revenue, Zhu2009Optimizing} & Learning to rank, logistic regression & 10m queries, 3 ads and search results for each query\\ \hline

Balance relevance and revenue & \citet{Radlinski2008Optimizing} & Regression SVM & 10m queries, 3 ads for each query, 5.4k judgements for substitution and 4k for matching ads\\ \hline
	\end{tabularx}
}
\end{table}

\subsection{Revenue Optimization}
\label{subsec-revenue}

As an ad exchange, the revenue optimization is among the most important tasks in order to run a business. This is especially true for search engine companies, for example, Google made 96\% of its revenue from its advertising business in 2010, 2011, and 2012 Q1\footnote{\url{http://investor.google.com/financial/tables.html} (last visited 04/06/2012)}, which supports the free, high-quality and fast expanding Google search engine. Without the profit made from sponsored search ads, all major free search services may no longer exist. Therefore, in the research area of revenue making there is a rich literature, mostly from industrial companies with big datasets.

With the GSP auction model and relevance restriction, the revenue seems fixed. Researchers generally follow two approaches to increase profit: 1) modify the GSP auction model, for example introducing the quality score to alter the ranking order 2) lower the relevance, or change the matching model to get more high-bid ads. These approaches will be shown in this section.

While people may worry that revenue optimization oriented ads ranking could drag down the CTR for sponsored links, \citet{Jansen2007Sponsored} reported from their experiments that 1) the CTR on all sponsored links combined is only around 15\%, much lower than the previously reported 35\% 2) people will retain their behaviour of clicking on organic or sponsored links even when these two types of result are mixed. This leads to a belief that from the user's perspective, the only thing that matters is the relevance of ads. Therefore it makes sense for ad networks/exchanges to try to increase revenue by re-ranking ads, as long as the relevance is not harmed.

\citet{Radlinski2008Optimizing} proposed a method of mapping queries to ads in an offline stage, and substituting new queries to existing ones in an online stage. The substitutions were selected based on the relevance score and second highest bid amount of any ad bidding on the candidate substitution. The experiment reported 2-3 times more revenue than the optimized method generated over the baseline (directly matching ads with queries).

If the GSP auction model is less weighted, the ad ranking problem focuses instead on finding a good balance between relevance and revenue. Following this idea, some researchers proposed models for obtaining new ranking functions. \citet{Mehta2005AdWords} named it the \textit{AdWords problem}, where a daily budget $b_i$ was restricted for $N$ bidders. Letting $Q$ denote the set of query words; each bidder $i$ specified a bid $c_{iq}$ for query word $q \in Q$. Then, when a sequence $q_0, q_1, \ldots, q_M$ of query words $q_j \in Q$ arrived online, each query $q_j$ must be assigned to some bidder $i$ for a bid of $c_{ij} = c_{i_{q_j}}$. The goal was to maximize the total revenue whilst respecting the daily budgets of the bidders.

 \citet{Zhu2009Revenue, Zhu2009Optimizing} proposed two machine learning algorithms trained using impressions and the CTR of real ads. They used three months of CTR log data (including impressions and ad clicks) to train a ranking function as described by \citet{joachims2002optimizing}. The function was limited by a predetermined value of revenue and tried to find the maximum relevance score. The empirical experiments showed that with the \textit{revenue direct-optimization} approach they were able to increase the revenue by 19.7\% (with 9 features) or 7.3\% (with 12 features) whilst avoiding losing relevance accuracy.

\citet{bala:ad_ranking_webDB08} introduced several elements of mutual influences between ads based on the typical page browsing patterns of users: 1) positional bias, usually the higher the position the more user attention an ad gets 2) similar ad fatigue, the perceived relevance of an ad is reduced by the presence of similar ads higher up in the list 3) browsing abandonment, the user may skip browsing the rest ads at any point of time depending on the ads already viewed. These elements will certainly affect the revenue, which when estimated, usually only take into consideration the position bias. The authors proposed a profit function with the optimality proved, considering 4) the relevance of ads as well.

\section{The Perspective of an Advertiser}
\label{sec-advertiser} 

Advertisers use ads to generate demand from potential customers for a product or service. Additionally, ads can be used to improve brand awareness, usually by displaying ads that are charged per impression. Advertisers can directly approach publishers to display ads on their webpages, and if accepted, the publisher will formalize a contract to guarantee the delivery of the requested number of impressions. Alternatively, by utilising the service of an ad exchange, an advertiser has the opportunity to have their ads displayed on the websites of many publishers; in this case, publishers sell their remnant inventory to the ad exchange \citep{roels2009dynamic}. More details concerning advertisers negotiating display contracts can be found in Section-\ref{sec-publisher}. 

In this section, we aim to describe the advertisers' role in Internet advertising and the key challenges faced by an advertiser using CPC pricing model, where keyword selection and bid optimization remain the most critical tasks. In the case of sponsored search, keywords are used to match which ads are relevant with the query terms issued by a user expressing their information need. Bidding is performed without knowledge of the bids of other advertisers for the same keywords, therefore, the bid should be a trade-off between the position offered and the profit could be gained provided a position. Because ad exchanges will select ads based on various factors including bidding price, an advertiser's bidding strategy is very important. Additionally, bidding for all possible keywords can incur significant costs, therefore research into designing experiments to select profitable keywords is also important. These challenges are discussed in the following subsections. 

\subsection{Keyword Selection and Bid Optimization} 

An advertiser aims to bid for keywords that best represent their products or services, however, the price of auction-driven keywords can vary. Generally, an advertiser aims to maximize their ROI by viewing each impression or click as an asset with future returns. In sponsored search, advertisers participates in keyword auctions, where they must select keywords before auctions start. For contextual or branding ads, advertiser negotiate a contract with publishers directly on the guaranteed delivery of ads, or via ad exchanges that act as intermediaries between advertisers and publishers. Recently, researchers have begun to investigate ways to improve advertisers' experience in the search-based advertising eco-system, modelling the keyword selection as an optimization problem constrained by the budget. 

The work of \citet{even2009bid} highlights the bid optimization problem when bidding on a subset of keywords in order to maximize profits. With all keywords biddable, they developed a linear programming based algorithm that runs in polynomial time that can gain optimal profit. When only a subset of keywords are biddable, the LP-based algorithm is no longer reliable, thus to remedy this problem constant factor approximation is used when the profit significantly exceeded the cost, based on rounding a natural LP formulation of the problem. In the paper, they study a budgeted variant of the bid optimization problem, and show that with the query language model one can find two budget constrained ad campaigns in polynomial that implement the optimal bidding strategy. Their results are the first to address bid optimization with \textit{broad match}, which is common in ad auctions. 
The work of \citet{rusm:keyword_ec06} focuses on selecting profitable keywords. As each keyword has different statistics, carefully selected keywords can lead to increased profits. The keyword statistics change from time to time, making keyword selection a daunting task. In their work, they developed an adaptive algorithm to select keywords for sponsored search, their algorithm prioritizes prefix ordering based on profit-to-cost ratio. The prefix ordering is ordered in descending order, where the top listed keywords are regarded as the most profitable. Although the number of keywords can number in the millions, their algorithm is believed to scale gracefully with the increase in keywords. Their simulations show that their algorithm performs better than the existing multi-armed bandit algorithms UCB1 and EXP3, measured by the increase in profits of 7\%. 

In their work on budget constrained bidding in keyword auctions, \citet{zhou2008budget} consider bidding optimization as a multiple-choice knapsack problem, defined as the following: suppose that there are $N$ available positions (slots), fix a keyword with position 1 through $N$. At time $t$, $X(t)$ is the number of clicks accumulated, $b_i(t)$ is the maximum bid for position $i$ and $V$ is the expected revenue. Then, winning at position $i$ at time $t$ causes the advertiser to be charged $w_i(t)$ and to gain a profit of $v_i(t)$. Assuming that $\alpha(i)$ is CTR, the cost and profit are expressed as follows: 
\begin{align*}
	w_i(t)&\equiv b_i(t)X(t)\alpha(i) \\ 
	v_i(t) &\equiv (V-b_i(t))X(t)\alpha(i) 
\end{align*}

\citet{feldman2007budget}  proposed a keyword bidding technique whereby they applied uniform bidding across all keywords; they claimed that the strategy was able to generate at least $1 - \frac{1}{\epsilon}$ fraction of the maximum clicks possible. Other significant work in keyword selection and bid optimization can be found in \citep{szymanski:ROI_wssa06, asdemir:dynamic_wssa06}. 

Some common metrics that can be used to measure advertising effects are clicks \citep{feldman2007budget}, revenues, conversions and clickthroughs. In budget constrained bidding, an advertiser specifies daily budgets when bidding for keywords as it is important for the advertiser to deliver marketing messages in a cost effective manner. Because ads are viewed as perishable items, one of the criteria that ad exchanges use to help determine which advertisers win a display slot is if the advertiser has a positive budget. In a typical ad campaign, an advertiser would have many ads, each having multiple associated keywords. With the assumption that all ads have future returns, all keywords associated with the ads potentially contribute towards the overall number of clicks. With a lot of ads to manage, setting a bid for each keyword can be a daunting task. 
\citet{zainal:adkdd10} attempted to optimize bids by maximizing the overall clicks for budget-constrained sponsored search ads. Given an overall budget and an average cost per click that the advertiser was willing to spend, they attempted to maximize the overall clicks using integer programming. By specifying the average cost per click, their method was able to take more risk by bidding higher on certain key phrases so long as the budget was well utilised and that the constraint on average CPC was met. They compared their method with a greedy strategy that maximized clicks, and their simulations showed promising results. 

\subsection{Experimental Design} 

As highlighted in the previous subsection, keyword selection is very crucial in any ad campaign, however, selecting all available keywords is not practical as it involves significant costs. A practical approach is to filter keywords that are historically profitable and bid on those keywords, although having a problem that some potentially profitable candidates are left unexplored, especially when they have no historical data at all. A multi-armed bandit strategy for keyword selection is an efficient way to trade off between exploitation and exploration. The term ``multi-armed bandit problem'' is a metaphor for a series of slot machines in a casino (sometimes referred to as one-armed bandits), where a gambler have $N$ slot machines to choose from, and at every time step he has to choose one to play at a time so as to maximise the return.  

Reforming the metaphor in the context of keyword selection, in the beginning $t=1$ we have no knowledge of how well a keyword attracts traffic. Therefore, the expected CTR for new keywords may have larger variability (or risk) than keywords that have already been used in the campaign, thus for short term performing an exploitation of known popular keywords might be useful. Alternatively an exploration should be performed to discover potentially profitable keywords and in the long run gain more from better candidates. Traditional multi-armed bandit approaches can be found in \citep{auer:finite_ml02}. Moreover, in the context of online advertising, ads also have features (context) that can be exploited to further refine retrieval tasks, which is commonly referred to as \textit{linear bandits} \citep{abernethy2008competing}.

Some researchers have started to make use of exploration and exploitation using the multi-armed bandit approach in the online advertising problem. In their paper, \citet{li:e_e_kdd10} develop methods that can adaptively balance the two aspects of exploitation and exploration by learning the optimal trade-off of the two and incorporating some historical performance confidence metrics that they developed. Their evaluation was based on log data and their algorithm performed quite well in terms of ad reach and CTR. 

Another practical concern is on the quality of ads. For example, if an ad is able to attract clicks but scores less conversions, an advertiser might be better off improving the landing page (or modifying ad creative for a better consistency with the landing page) rather than their keyword selection. Running an experiment for all of different feature possibilities comprising an ad can be costly, therefore, a Taguchi method \citep{ranjit:taguchi, taguchi1987system} to run the experiment can significantly save a lot of effort by reducing the number of trial instances needed.

\section{The Perspective of a Publisher}
\label{sec-publisher} 

Most publishers (e.g. CNN, BBC) reserve some space in their website for branding ads or contextual advertising. Inventories can be sold in the form of contracts or in real-time. The publisher is responsible for delivering a total number of impressions based on what was agreed on in the contract. In the event that the publisher is unable to deliver all guaranteed impressions, a penalty applies. Therefore a challenge for publishers is to select the optimal contract or estimate the optimal price.

In contextual advertising where the revenue is being shared between publishers and ad exchanges, the strategy for publishers is to have content that contains popular and profitable keywords. This is essential because the keywords found by the contextual advertising system are matched with the keywords selected by advertisers who are willing to have their ads displayed on publishers' websites. One of the key challenges for a publisher is to balance between revenue (and relevant ad displays) and the quality of content, the more relevant the display ads, the more likely users will click and the more revenue that can be earned. With the CPC model, if impressions do not convert to clicks, then the publisher is actually promoting an advertiser's ads for free. 
 
The ultimate challenge for publishers is to maximize their revenue. Referring back to the cash-flow in the advertising ecosystem, a user is able to browse a website for free because the service is \textit{supported} by the publisher, whose revenue is partly generated from selling display space to advertisers; revenue maximization is therefore crucial. 

\subsection{Revenue Maximization} 

In the domain of search advertising (and applicable to contextual advertising as well), relevant ad matching has been carried out by \citet{Mehta2005AdWords} using a multi-armed bandit formulation \citep{pandey2007handling}. In contextual advertising, the revenue is shared with an ad exchange, how and when an ad is delivered in the publisher's website is not within the publishers' control. The contents of the ad play a vital role in determining the relevance of an ad within a webpage. 

The focus here is on maximizing the revenue from the OTC contracts; the strategy here is more apparent because the publishers have more control on the advertising requests that they receive directly from advertisers. In order to maximize revenue from the OTC contracts, the publisher's allocation and inventory management have to be efficient. The work by \citet{roels2009dynamic} and \citet{feige2008combinatorial} incorporated contract guarantees, where an attempt was made to maximize a publisher's revenue in display ads through dynamic optimization. Their model allows publishers to dynamically select which advertising requests are awarded advertising contracts. Each advertising contract $i$ consists of a set of contracts ($n_i$, $\mathcal{W}_i$, $r_i$, $\pi_i$) where $n_i$ is the number of impressions requested, $\mathcal{W}$ is the set of webpages/viewers/time of delivery, $r_i$ is the cost-per-thousand (CPM) impressions and $\pi_i$ is the goodwill penalty in case the requested number of impressions are not fulfilled. 

Unlike a store that holds physical items, the total number of ad impressions is generally unknown a priori as web traffic plays a role in estimating the number of available impressions. \citet{roels2009dynamic} argue that unlike broadcast networks, Web publishers are able to observe the traffic to their websites and therefore can dynamically plan a strategy to deliver impressions. In their dynamic programming formulation to maximize revenue, they balance the immediate revenues at time $t$ against the potential future profits. This strategy is useful to decide whether or not to accept incoming advertising requests at time $t$ when there are existing contracts that are still valid, when the difference between the number of requested impressions and the number of delivered impressions up until time $t$ is positive. Dynamic programming solutions can be intractable for realistic problem instances, so the authors present a Certainty Equivalent Control (CEC) heuristic that reduces the complexity of the formulation to the integer programming level and is more feasible for a real case study. The performance of their optimization approach has been evaluated on a real problem instance using Break.com, a website that serves display ads on its homepage, video page, game page and pages on its website as a case study. Using CEC, the increase in profits was 15-20\% more than using Break.com's old strategy to accept new advertising requests. 

\subsection{Scheduling and Improving Content Quality} 

Scheduling an ad display is also crucial for Web publishers. \citet{nakamura:LP_ecr05} developed an LP-based algorithm to schedule banner ads, where they presented three features that each ad was associated with; time of day that the ads were preferred to be viewed (e.g. the afternoon), page category (e.g. sports) and the number of impressions. The features were then used to determine the optimal ad time and location that maximises overall revenue, rather than relying solely on the CTR of an individual ad. Their strategy showed an improvement over greedy and random methods. 

In order to attract users to visit a website and eventually to click on the ads listed in the website, the content and the page presentation must be interesting to navigate. A web page consists of many features, to test all combinations of features is costly, particularly when displaying them on a live page. To illustrate, suppose that we would like to design a web page with a good combination of background and text colours, some combinations are better in terms of web page readability, and there are 32 different colours to choose from. To test each combination requires $32\times32 = 1024$ experiments, where each experiment requires several subjects. Several approaches are possible; the first choice is to go through all background colours and choose one, try all possible text colours and view the best combination. The second option is to choose the text colour first and then try several background colours and view the best combination. The third option is to set up a criterion for readability, and later select a few pre-selected colours for the background, then try some text colours to go against the chosen background colours. The first text colour choice meeting the requirement is then the selected combination. 

The problem with the second and third options is that several choices of background/text colours may not get a number of reasonable trials, despite potentially yielding a good design. A fractional factorial design \citep{gunst2009fractional} is applicable in this case to avoid redundancy in the number of trials. The main principle in factorial design is orthogonality, assume that we have two feature vectors $\mathbf{A} = (a_1,\ldots,a_n)$ and $\mathbf{B} = (b_1,\ldots,b_m)$, the set of experiments is orthogonal if each pair-wise combination of values $(a_i,b_j)$ occurs in the same number of trials. 

Once the factorial experiments have been conducted, the results are analysed to select the best design. The Taguchi method \citep{ranjit:taguchi, taguchi1987system} suggests that the \textit{quality} should be thought of, not as a product of adhering to a set of specifications, but by the variation from the target, where the different designs are compared in terms of their signal-to-noise ratio. 

\section{The Perspective of a User}
\label{sec-user} 

The user is a person who browses the web, consumes media content and performs searches using search engines. In the process of performing these actions, the user is exposed to online advertising (in its many forms) and may or may not acknowledge, view or engage with ads. The user is central to any web advertising campaign, publishers wish to attract users to their content so that they are exposed to the ads that help finance the website, advertisers attempt to entice users into clicking on their ads by making them contextual and appealing, and search engines aim to provide ads that are relevant as well as profitable alongside their search results. Users seem to acknowledge that the presence of ads is what allows them to view much of the content on the web for free and have accepted web advertising as a way of life on the Internet \citep{mcdonald:perceive_advert_ecir09}. 

Nonetheless, \citet{enquiro:barriers08} demonstrated that users show a reluctance to click on web advertisements and make an active effort to avoid doing so, viewing advertising as a visual barrier obstructing the content of the page. To further illustrate this point, \citet{jansen:non_sponsored06} showed that users performing e-commerce searches were more likely to click on a non-sponsored link for a website than the same link contained in an advert. It seems evident that whilst users recognize the need for web adverts, they tend to behave as if they would prefer them not to be there. 

This serves to highlight one of the distinguishing differences between web advertising and traditional advertising, the fact that users have interactive freedom whilst browsing the web. Watching television tends to be a fairly passive activity with viewers often willing to continue watching during advertisement breaks, whilst print ads are more effective at capturing reader attention than similar web ads \citep{gallagherEffectiveAds}. Whilst accessing the web, users are active participants in information finding and thus are quick to disregard information that is not relevant; the interactive nature of the Internet allows users to opt out of engaging with advertising. Thus, it is vital that in order to effectively advertise on the web, advertisers need to discard many traditional techniques and find new ways of appealing to the user, who now has access to a rich amount of information but only a limited amount of attention \citep{huberman2007economics}. 

In response, advertisers can make use of the context of user actions and their personal information in order to tailor adverts to their demographic and taste, so as to maximise the number of click-throughs received for each ad and subsequently the revenue obtained. But, before exploring further how user context and personal information can be used to target ads, we'll first consider the role that the user plays in responding to ads and how this affects what ads an ad exchange (or publisher) will show to users. 

\subsection{Click-throughs} 

\begin{table}[t!] 
\small{ 
	\centering 
	\caption{Categorisation of eye tracking research, user browsing models and other clickthrough related research} \label{tbl-user-cat} 
	\begin{tabularx}{\textwidth}{|X|X | X | X|} \hline
\rowcolor{Gray}
	Goal & Author(s) & Method & Data \\ \hline 

\multirow{5}{*}{Learn query intent} & \citet{ashkan2009classifying} & SVMs, Kernel Methods and click-throughs& 135,000 commercial search engine queries and ad data\\ \cline{2-4}

& \citet{ashkan:query_intent_click08} & Decision Trees and clickthroughs & 135,000 commercial search engine queries and ad data\\ \hline 

Query Rewriting& \citet{Zhang:Rewriting} & Clickthroughs with Logistic Regression and editDist/wordDist& 2000 queries from web search log\\ \hline 

\multirow{5}{*}{\specialcell{Estimating CTR \\ for new ads}}& \citet{regelson:keyword_clusters06} & Keyword clustering and smoothing& Commercial search logs and ad data\\ \cline{2-4}

& \citet{Richardson2007Predicting} & Click-throughs with Logistic Regression, feature extraction& Commercial search logs and 1 million ads\\ \hline 

\multirow{2}{*}{\specialcell{Accuracy of clicks \\ as implicit \\ judgement}} & \citet{granka2004eye} & Eye tracking and survey & 26 participants\\ \cline{2-4}

 & \citet{joachims:ctr_ir05} & Eye tracking and survey& 29 participants\\ \hline 

What acts as a barrier on search results page & \citet{enquiro:barriers08} & Eye tracking & Experimental\\ \hline 

Google eye tracking& \citet{enquiro:eye_tracking05} & Eye tracking & 48 participants\\ \hline 
	\end{tabularx} 
} 
\end{table} 

Following careful keyword analysis, successful bidding for an advertising position and potential targeting of adverts to a user, what is the best way of determining if an online advertisement is in fact working? Explicit techniques, such as asking users or experts to evaluate the relevance/quality of an advert are too expensive or intrusive. \citet{ Fox:Implicit_measures} found that implicit measurements, in particular click-throughs, time spent on page and exit type (closed browser, new query session etc.), observed whilst users performed web searches and read news articles, correlated with explicit user satisfaction measurements and thus could be used as an approval metric. 

Nowadays, binary click-through flags (whether a url has been clicked or not) are the primary method for capturing a user response and determining user feedback and interest in web advertising, and increasingly used in web advertising revenue models due to their simplicity, lack of intrusion and availability. The list of pricing models available to an advertiser include \textit{CPA, CPC, CPM, CPV} and etc. Many of the models involve the user actively interacting with an advert (usually by clicking on it) before the advert is paid for, thus there is active research in the area of how best to predict the \textit{CTR} for particular adverts. 

Unfortunately, click-through data is inherently noisy; whilst navigating search and content web pages, users may click indiscriminately, irrationally, malevolently (for example \textit{click fraud}, the act of paying people or creating algorithms that actively click on ads so as to create revenue for a publisher or increase the cost of a rival advertisers campaign) or misguidedly \citep{agichtein2006learning}, as well as positional bias (covered in more detail in the next section). Nonetheless, it is the abundance of click-through data that allows it to be usable, the correct analysis of which has allowed search engines to successfully infer the general behaviour of users and optimise their search results and ad placements. 

Click-through data, whilst abundant, can be notoriously inaccurate for advertising, as quoted by \citet{Richardson2007Predicting}, ``For example, an ad with a true CTR of 5\% must be shown 1000 times before we are even 85\% confident that our estimate is within 1\% of the true CTR'' and ``In general search advertising, the average CTR for an ad is estimated to be as low as 2.6\%''. Typically, the CTR of an ad is calculated using the Maximum Likelihood Estimator: 
\[ 
	\frac{\sharp(\mbox{number of clicks})}{\sharp(\mbox{number of impressions})} 
\] 
where an \textit{impression} is a displayed ad. Due to the low rate that is observed for ads, this metric suffers from a high variance and can be misleading. An additional problem is that for new ads with few or zero impressions, it can be difficult to adequately predict the click-through rate and thus maximise ad placement revenue. 

For new ads, these limitations can be overcome by employing additional evaluation metrics. \citet{Richardson2007Predicting,  kushal:rare_new_ads10} and \citet{ regelson:keyword_clusters06} proposed using machine learning algorithms to analyse a new ad's content and quality so as to cluster them together with semantically similar (using keywords and extracted terms) existing ads, and then predict the CTR by using historical CTRs. Similarly, \citet{Zhang:Rewriting} suggests using click-throughs as an implicit measure for evaluating the success of rewriting search queries so as to improve user spelling, or find synonyms of ad targeted keywords so as to better target related ads. More recently, \citet{guo2010ready} proposed inferring user intent (categorised into research and purchase) from user mouse behaviour (such as cursor movement and scrolling), as well as click-throughs, and using this information to determine the correct search results to display, as well as whether or not to show ads to the user. 

\subsubsection{Positional Bias} 

A previously mentioned disadvantage of using click-throughs as an implicit feedback mechanism was that they can be affected by positional (or rank) bias. This effect was confirmed by \citet{joachims:ctr_ir05} whilst using an eye tracking device to record users' attention whilst navigating a search results page. In the study, it was found that there was a tendency for lower ranked search results to receive less attention and thus less click-throughs than those of a higher rank (which is also applicable to lists of adverts).
The study concluded that click-through data was synonymous with document relevance, and that users generally scanned web pages from top to bottom with decreasing interest.


A further commercial study \citep{enquiro:eye_tracking05} used eye tracking to explore how users navigated the Google search interface and revealed much about how sponsored adverts on the search page were regarded by the user. It was found that by introducing sponsored search results above the \textit{organic} listing, that the users attention would subsequently be drawn to the sponsored listings. In addition, a significant number of users would actively scan or read the sponsored listing, attracting a sizeable volume of click-throughs and thus ad exposure. This proved significant when compared to side sponsored ads, which were looked at by fewer users and clicked less often. However, when returning to a results page for the second time, users tended to be less discriminative and more likely to look at and click on a side sponsored advert than a top sponsored ad. 

Another finding of the study was that when users did click on one of the top sponsored ads, that they managed to find what they were searching for a higher proportion of the time than when clicking on organic ads. This result was also found by \citet{jansen:comparitive_effectiveness07} when using solely e-commerce style queries, giving an indication that search engines are adept at providing relevant ads for users to click on, which helps in building consumer trust in the reliability of such ads. 

\subsection{Behavioural Targeting} 

Related to the field of contextual advertising is that of \emph{Behavioural Targeting}, where instead of displaying ads that are relevant to a given query or the context of a webpage, ads are made more relevant by tailoring them specifically to the user themselves. An early example of behavioural targeting is when inferring the intent of a search in a search engine, and displaying ads accordingly \citep{jansen2008determining}. \citet{ashkan2009classifying} classes users into categories performing either \textit{commercial}/\textit{non-commercial} and \textit{informational}/\textit{navigational} searches, with each class having different responsiveness to search advertising, and where commercial-navigational queries garnered the most ad click-throughs. 

More recently, websites have made use of an array of \textit{Personally Identifiable Information (PII)} about users whilst they are accessing the website, derived from cookies, flash cookies, web beacons, browser and other meta-data. This PII is used to profile users and so deliver relevant, targeted ads to them, which has been shown to be effective by \citet{yan:behavioural_helps_advertising} and \cite{jaworska2008behavioural}, who were able to provide empirical evidence of improvement using behavioural targeting. Even in the absence of cookies and other tracking files, it is possible to perform user profiling based on browser data alone \citep{ eckersley2010unique}. 

Once a user can be uniquely identified, then they can be tracked across the pages of a publishers web site, or across the websites comprising the same ad exchange or data exchange. The pages that the user visits are recorded using cookies and can be used to infer to some degree of accuracy additional demographic information (such as age, gender) about the user \citep{hu2007demographic}, as well as the users interests. Users can then be clustered into profile groups, which can be targeted specifically by advertisers \citep{ Wu:Semantic_segmentation}, for instance, demographics such as gender and age or interests such as sports or fashion. 

PII is not just restricted to the aforementioned sources; as technology improves, so too do opportunities to gather more reliable, contextual and intrusive information about a user. The dramatic increase in mobile device web access in recent years has led to the development of sophisticated geo-location technology and \textit{geo-targeted ads}. Another example is the analysis of email content to inform advertisements, which is currently performed by Gmail. Furthermore, the shift towards social media on the web has led to an influx of users willing to store revealing personal information online as well as information about their relationships to other people, information that allows for better ad targeting. In addition, such websites are also shifting the way that users engage with adverts, allowing users to become a fan of advertisers and use the social network itself to disperse ads through friend recommendations. As such, advertiser spending in social media has dramatically increased, although as \citet{Webtrends_facebook} shows there are still more lessons to learn in how to effectively make the most of the network. 

Another source of user information can be provided by ISPs and is known as \textit{Deep Packet Inspection}, although this technique has been plagued with controversy and so has not been widely implemented. The most notable examples of its use were by NebuAid, who were shut down after US Congress cracked down on DPI techniques \citep{NebuAid_US}, and Phorm\footnote{\url{http://www.phorm.com/} (last visited 13/12/2011)}, who caused controversy in the UK when ISP BT started trialling Phorm's DPI tools on users without informing them \citep{Phorm_UK}. 

\subsubsection{User Privacy} 

\begin{table}[t!] 
\small{ 
	\centering 
	\caption{Categorisation of behavioural targeting and user privacy research} \label{tbl-user-policy} 
	\begin{tabularx}{\textwidth}{|X|X | X | X|} \hline 
\rowcolor{Gray}
	Goal & Author(s) & Method & Data \\ \hline 

User policy Predicting demographic information& \citet{hu2007demographic} & SVM, Singular Value Decomposition & 190,000 click-through logs for website\\ \hline 

Behavioural Targeting overview& \citet{jaworska2008behavioural} & Decision Trees, Naive Bayes, Discriminant Analysis & 1.5 million ad impressions on commercial websites\\ \hline 

Behavioural Targeting case study& \citet{dwyer_levis_case_study} & Analysis of Levis.com website & Levis.com website\\ \hline 

User segmentation& \citet{Wu:Semantic_segmentation} & Probabilistic latent semantic analysis & 1 day's commercial search engine ad data\\ \hline 

The perceived relevance of sponsored links& \citet{jansen:non_sponsored06} & Web search observation and survey & 56 participants\\ \hline 

\multirow{2}{*}{\specialcell{User perception of \\ behavioural \\ targeting}} & \citet{mcdonald:perceive_advert_ecir09} & Interview survey & 14 participants\\ \cline{2-4} 

& \citet{truste:consumer_09} & Survey & 1,015 participants\\ \hline 
	\end{tabularx} 
} 
\end{table} 

The issues surrounding deep packet inspection are just one of a number of problems that behavioural targeting technologies have had to contend with. In the case of DPI, there exist campaign groups such as NoDPI who have repeatedly sued ISPs over privacy issues. In addition, other forms of contextual advertising have also come under fire, including Gmail \citep{bbc_gmail}, Facebook \citep{rosenblum:privacy_social_07, mashable_fb_beacon} and Google's predecessor DoubleClick \citep{cnet_doubleclick}. 

Users are discomforted by the knowledge that their online actions are observed and recorded (as observed by \citet{truste:consumer_09}) and are often clueless as to what tools are out there that can help protect their privacy. Currently, much of the advertising industry is self-regulated and follows the guidelines set out by the \cite{self_regulatory:09}, with self regulation argued as the ideal solution so that the web can remain free and accessible to all, and so that subscription based models can be avoided (which \citet{szoka:targeted_online:09} argues as being harmful to the ability of websites to innovate and publish creative content). Nonetheless, it does appear that governments have taken more notice and in some places such as the UK new regulations regarding online advertising are starting to appear. Moreover, many websites who do make use of behavioural targeting are attempting to clarify their policies and allow their users to take a more pro-active role in managing their privacy. \citet{ dwyer_levis_case_study} provides a good overview of the behavioural targeting process for one website, and in particular, illustrates the inconsistency between website privacy policies and their actual policies. 

Privacy is still a concern among users and there exists an increasing number of ad-averse web users who make efforts to avoid being tracked by ad networks. Browser plug-ins such as AdBlock Plus \footnote{\url{https://addons.mozilla.org/en-us/firefox/addon/adblock-plus/}  (last visited 13/12/2011)} and BetterPrivacy\footnote{\url{https://addons.mozilla.org/en-US/firefox/addon/betterprivacy/}  (last visited 13/12/2011)} for FireFox allow users to better control what websites can access their data, and websites exist that allow users to determine their PII identifiability\footnote{\url{https://panopticlick.eff.org/}  (last visited 13/12/2011)}. In addition, new advertising platforms, such as Adnostic \citep{ Toubiana_adnostic:privacy}, aim to deliver targeted ads but without allowing user information to leave their computer. 

\section{Conclusion and Future Directions of Internet Advertising}
\label{sec-future}

In conclusion, this paper highlights the technology in Internet advertising which has become popular in recent years. From the perspectives of advertisers, online publishers, ad exchanges and web users, we have presented the brief history and the overview of the entire ad eco systems and business models, and analysed and compared the technical challenges and recent solutions.

In future, it is anticipated that ads delivered to users will become more targeted, where all participants in the eco-system are harmonized by increased utility and satisfaction. Advertisers strive towards advertising whose costs are minimized whilst ROI is maximized; the mechanisms to achieve these goals are (i) to have a system able to select keywords that matches search queries precisely; (ii) have a good algorithm of learning and predicting user behaviour based on browsing history; (iii) have a more advanced strategy of bidding. 

For publishers, the management of allocation and inventory is crucial, it is therefore important for publishers to have a trade-off between selling their spaces based on contracts and auctions. The inventory sold through contract gives guarantees of publisher revenue, nevertheless, the auction-driven inventory makes good utility of remnant impressions and can bolster income occasionally. In order to improve the quality of the content/page, perhaps fractional factorial design could be employed to test which features of the websites are attractive to users, which could also be employed by advertisers to improve the quality and ROI for ads as well. 

For ad exchanges, the challenges lie on maintaining a healthy eco-system by improving the auction mechanism, providing more insight for both advertisers and publishers, and identification and prevention of fraud. It seems promising that the whole industry is moving towards seamless integration of data from different marketplaces, resulting in a uniform and economically efficient market. The current web advertising is generally in the hands of a few key players. Google effectively maintains about 48\% of the search-based advertising market, while the rest is shared mainly by Yahoo! and Microsoft, according to a recent survey from Financial Times. The consequence is that two display opportunities with similar targeted audiences and visit frequency may sell for quite different prices on two different markets. An advertiser may pay a higher price for similar display opportunities, which are otherwise sold with a much cheaper price from another market. Equally, if you undertake a campaign to sell travel insurance for instance, investing in few highly visited web pages might be more costly than consolidating the same quality display opportunities from a wide range of unpopular personal blogs about travels. We thus need to establish a unified ad exchange to unify the effectiveness measure so that display opportunities can be widely accessible, fairly compared and traded, benefiting all the participants in the long run. Research efforts are required to create a unified ads exchange that creates a neutral computer-mediated clearinghouse to enable publishers (ad providers) to fairly get maximum yield from their provided display opportunities, while enabling advertisers (ad consumers) to efficiently find all relevant display opportunities with complete transparency regarding quality and value, evaluated against their own buy criteria. 

Moreover, current ad exchanges are limited mainly to ��spots�� markets, i.e., any transaction where delivery takes place immediately. There are no technologies available to support efficient forward pricing and risk management mechanisms for web advertising, which are important to business. It is an urgent need to develop techniques to equip the spots market with ad forward or standardised futures contracts (agreements to buy or sell display opportunities at a certain future time for a certain price) capabilities \citep{adoptions}. The following example illustrates how important the futures contract capability is to web advertising. Suppose there is a travel insurance company whose major customers are found in the web. In March the company plans an ad campaign in three months time as they think there will be more opportunities to sell their travel insurance products in the summer. Display opportunities are usually priced on the basis of the supply and demand, and their prices influence the business decision. If the company worries that the future price of the display opportunities will go up, they could hedge the risk (lock in the campaign cost) by agreeing to buy the display opportunities in 3 months time for an agreed price (taking a long position in a 3-month forwarding market). Equally, search engines and large content publishers could agree to sell display opportunities in the future (take short positions on futures contracts of their display opportunities) if concerned that the price will go down and enabling them to lock in a profit. Intuitively, ad Futures Contracts are also useful for inventory management. For instance, if a publisher has agreed to deliver a certain amount of display opportunities (usually measured by impressions, appearing times) to some advertisers in one month time, but now predicts that there will be no adequate impressions (visits) available from its own inventory (available web pages), what they can do is to take a long position, agreeing to buy more impressions from the futures market if the market��s futures price is currently lower or equal.  \citet{adoptions} argues that affected by the dual force of supply and demand, the prices of the ad slots vary significantly over time. The fluctuated ads prices make the costs for advertising rather unstable and risky, and the plan of ad campaigns difficult. To address the issue, the authors proposed to sell future's ad impressions via option contracts and give the advertisers right, but no obligation, to buy futures ad slots at a fixed price.

For users, behavioural targeting is a recent trend. It is an emerging field that is still trying to find a healthy balance between ultimately exploiting user information and respecting privacy. Improvements in the fields of unsupervised learning, clustering and de-anonymization will allow advertisers to infer a greater level of detail about users and consequently better targeted ads, although this technology will have to compete with increasing numbers of users who take active measures to protect their privacy. In addition, new technology, such as the prevalence of location aware smart phones and web connected games consoles, provide new sources of information and present new privacy challenges. It has been reported that interactive billboards are currently under development and are capable of tailoring ads to the gender and age of passers by \citep{Gary2011Minority}. It seems inevitable that such ads could one day connect to a users mobile device and access advertiser accumulated target information.

One of the currently promising new technologies is social media, which as reported is redefining how users engage with ads and the amounts of PII that users are willing to volunteer. Nonetheless, there is little current research into how best to exploit such an abundance of extra information, and how best to take advantage of interactive ads, although there is an increasing amount of new research into using social media to predict trends, which could be useful for keyword bidding.

In addition, research on using click-through data is still very active due to the large quantities of such data available to advertisers and also the relevance in using this kind of data to improve search engine revenue. More accurate user browsing models may be developed, including those specific to ads with low click-through rates, and new techniques for using click-through as a feedback mechanism in a dynamic system, or as reward in a ranking multi-armed bandit.


\bibliographystyle{model4-names}
\bibliography{survey}


\appendix

\section{Terminology}
\label{sec-terminology}

\begin{description}
\item[Ad Exchange] A marketplace for advertisers to buy impressions and for publishers to sell them.

\item[Bidding Phrase] A keyword that advertisers bid for to represent their anticipation of the query that a user likely to submit.

\item[Clickthrough Rate (CTR)] A clickthrough occurs when a user clicks on a link or an ad. The CTR is the number of clicks over the number of impressions.

\item[Commercial Search] Searches that involve e-commerce in some form, i.e. online shopping.

\item[Conversion Rate] A conversion occurs when a user completes some desired action after reading the ad. The conversion rate is the number of conversions over the number of impressions.

\item[Cookie] A small file placed on the users computer by a website that is usually for saving personal information and could be used by advertisers for targeting.

\item[Cost per Action (CPA)] The price that the advertiser pays if the user accesses the website via the ad and performs some kind of action such as filling in a form, registering or purchasing.

\item[Cost per Click (CPC)] The price that the advertiser pays if the ad was clicked.

\item[Cost per Complete View of Video Ads (CPCV)] The price that the advertiser pays if at least a certain percentage of a video ad was played to user.

\item[Cost per Mille-Impressions (CPM)] The price that the advertiser pays if the ad was displayed to a user, using 1000 impressions as the unit for ease of presentation.

\item[Cost per Full-Page View (CPV) ] The price that the advertiser pays for every single display of the ad, which is usually in the form of pop-up or full-screen.

\item[Cost per Lead/Visit (CPL)] The price that the advertiser pays every time a targeted visitor appears on the advertisers website

\item[Creative] A brief description about a service or product that the advertiser wishes to promote.

\item[Data Exchange] A marketplace for ad exchanges to buy user profiles for better matching in real-time bidding.

\item[Deep Packet Inspection] Investigating the contents of web packets in real time in order to extract user information and Internet behaviour. Can be used by ISPs to prohibit user actions and censor information, and also by advertisers for targeting.

\item[Demand Side Platform] An automated bidding platform for advertisers to get good impressions at low cost, by participating in multiple auctions among various ad exchanges at the same time.

\item[Forward Contract] The non-standardized contracts showing the agreement of purchasing some goods in the future at the agreed price.

\item[Generalized Second Price Auction (GSP)] \citep{Edelman2007Internet} The advertiser pays the next highest bid instead of their own bid price.

\item[Informational Search] The classic Information Retrieval definition of a search, whereby users will use broad query terms to satisfy an information need, and may perform multiple searches using more refined queries.

\item[Impression] An impression occurs when the ad is displayed to any user.

\item[Landing Page] A web page associated with an ad, which will be shown to the user after the ad has been clicked.

\item[Navigational Search] A search where users have a particular destination in mind and are using a search engine as a conduit to that destination, i.e. searching for 'UCL' in order to gain access to the University College London website.

\item[Organic Search Result] Search results that are determined by search algorithms and are not paid for being displayed.

\item[Over-The-Counter (OTC)] \citep{downes1991dictionary} Non standardized products traded privately between two parties.

\item[Personally Identifiable Information (PII)] \citep{mccallister2010guide} Information about a user that can be determined from their cookies, browser information and IP address, amongst other sources.

\item[Pre-Set Bidding (PSB)] The advertisers have to specify beforehand the targets of an advertising campaign, including keywords, geography, language, devices, time, placement and so on, as well as the bids and daily budgets. When auctions are carried out advertisers cannot change the bids.

\item[Return-On-Investment (ROI)] \citep{downes1991dictionary} Usually expressed as a percentage, ROI is the ratio of money gained or lost (whether realized or unrealized) on an investment relative to the amount of money invested.

\item[Real-Time Bidding (RTB)] \citep{mike2009realtime} Similar to PSB, advertisers set parameters before running the campaign. However when auctions are carried out for every impression, additional data (e.g. context and user profiles) is passed to advertisers who can then change the bid accordingly for the specific impression.

\item[Supply Side Platform] An automated platform for publishers to sell impressions at an optimal price, by creating multiple auctions for the same impression in different ad exchanges to reach more advertisers willing to bid.

\item[Transparent Market (TM)] On the contrary with OTC, trading in TM are made in an open and public fashion so that buyers and sellers can usually see the attributes of other trading as a reference.

\item[User] A person who reads ads whilst engaging in some activity such as browsing the web.  

\end{description}

\end{document}